\documentclass[aps,prd,twocolumn,amssymb,nofootinbib,longbibliography]{revtex4-2}
\usepackage{ulem}
\usepackage[usenames,dvipsnames,svgnames]{xcolor}
\usepackage{color}
\usepackage{graphicx}
\usepackage{epsfig}
\usepackage{subfigure}
\usepackage{booktabs}
\usepackage{amsfonts,amsthm,mathrsfs,amsmath}
\usepackage{lipsum}
\usepackage{dutchcal}
\usepackage{times}
\usepackage{inputenc}
\usepackage{bm}
\usepackage{multirow}
\usepackage{url}
\usepackage{natbib}
\usepackage[colorlinks=true,citecolor=Blue,urlcolor=Blue,linkcolor=Blue]{hyperref}

\newcommand{\be}{\begin{equation}}
\newcommand{\ee}{\end{equation}}
\newcommand{\ba}{\begin{eqnarray}}
\newcommand{\ea}{\end{eqnarray}}

\def\That{{\hat{T}}}
\def\lambdahat{{\hat{\lambda}}}

\begin{document}
\title{Shear induced cavitation in radially expanding, chemically equilibrating QGP
}
\author{Lakshmi J. Naik}
\email{jn\_lakshmi@cb.students.amrita.edu}
\author{V. Sreekanth}
\email{v\_sreekanth@cb.amrita.edu}
\affiliation{
Department of Physics, Amrita School of Physical Sciences Coimbatore, Amrita Vishwa Vidyapeetham, India}
\date{\today}
\begin{abstract}

We present the first study of shear viscosity induced cavitation incorporating transverse expansion alongside chemical non-equilibrium dynamics in the Quark-Gluon Plasma (QGP). Chemical non-equilibrium is incorporated through quark and gluon fugacities, whose evolution is coupled to causal dissipative relativistic hydrodynamics in the Gubser geometry. Using phenomenological temperature-dependent shear viscosity parametrizations, we study the evolution of the longitudinal pressure and identify the conditions under which it becomes negative and thereby breaking down the hydrodynamic description. We demonstrate that shear induced cavitation develops first at the centre of the fireball ($r=0$) at early times and subsequently extends towards larger radial distances, with transverse expansion accelerating its onset, while persisting during chemical equilibration. Within our framework, we determine the critical initial shear viscosity $(\eta/s)_{\rm crit}$, below which cavitation does not occur throughout the QGP evolution. 
Since hydrodynamics is found to successfully describe the phenomenology of heavy-ion collisions, our results provide a constraint on shear viscosity requiring the absence of cavitation during the QGP evolution. We find that the phenomenologically extracted $\eta/s$ from JETSCAPE analyses lie below the critical shear viscosity $(\eta/s)_{\rm crit}$ obtained in our framework. 
\end{abstract}
\maketitle

Understanding the transport properties of the strongly coupled hot and dense Quark-Gluon Plasma (QGP) has been one of the central goals in relativistic heavy-ion collision studies over the past two decades~\cite{Meyer:2011gj,Braun-Munzinger:2015hba,Shuryak:2014zxa,Jaiswal:2020hvk}. Among these, the shear viscosity is one of the important transport coefficients as it plays a major role in governing the collective flow and final observable spectra measured in the experiments. Studies suggest that QGP possesses a small value of ratio of shear viscosity to entropy density $\eta/s$~\cite{Hirano:2005wx,Romatschke:2007mq,Heinz:2013th}, close to the KSS bound $1/4\pi$~\cite{Kovtun:2004de}. 
Shear viscosity, however, can exhibit a strong temperature dependence, with a minimum near the phase transition, as observed in other realistic fluids~\cite{Arnold:2003zc,Csernai:2006zz,Lacey:2006bc,Niemi:2011ix,Denicol:2015nhu,Vujanovic:2017psb}. 
\par
The temperature dependence of the shear viscosity of QGP is still non-conclusive and obtaining a consistent formulation of $\eta/s(T)$ remains an active research avenue. Considerable efforts have been devoted to calculate the temperature-dependent specific shear viscosity of quark-gluon matter. Several theoretical calculations of $\eta/s$ exist using various approaches~\cite{Arnold:2003zc,Meyer:2007ic,Fuini:2010xz,Chen:2009sm,Christiansen:2014ypa,Astrakhantsev:2017nrs,Borsanyi:2018srz,Liu:2016ysz,Ohanaka:2026hjx}. The temperature dependence of $\eta/s$ has also been extracted from phenomenological multi-stage frameworks which combine parameterizations of $\eta/s(T)$ with the modeling of initial conditions, viscous hydrodynamics, and hadronic transport~\cite{Shen:2011eg,Niemi:2012ry,Bernhard:2016tnd,Bernhard:2019bmu,Parkkila:2021tqq,JETSCAPE:2020mzn,Nijs:2020ors,Parkkila:2021yha,Heffernan:2023utr}. 
\par
Causal second-order relativistic viscous hydrodynamics has emerged as a standard framework to describe the evolution of QGP created in ultrarelativistic heavy-ion collisions, successfully explaining a broad range of collective flow observables~\cite{Luzum:2008cw,Romatschke:2017ejr,Busza:2018rrf}. The relativistic first-order Navier-Stokes theory~\cite{Landau:1987,Eckart:1940te} displays acausal behaviour~\cite{Hiscock:1983zz,Hiscock:1985zz} and this has led to the formulation of causal higher-order viscous hydrodynamic theories~\cite{Muronga:2001zk,Rocha:2023ilf}. The relativistic hydrodynamic equations are solved by taking the geometry of heavy-ion collisions to account. The simplest analytical model that successfully describes the geometry of QGP evolution is the one-dimensional boost-invariant Bj\"orken flow~\cite{Bjorken:1982qr}. A generalization of the same was developed by Steven Gubser, by considering the radial expansion together with the boost invariance, known as Gubser flow~\cite{Gubser:2010ze}. 
\par
The success of relativistic dissipative hydrodynamics in describing the collective behaviour of QGP relies on the assumption that the system remains close to the local thermodynamic equilibrium. However, strong viscous gradients present in the system may result in sufficiently large viscous pressures compared to the equilibrium value leading to the scenario of \textit{cavitation} and consequently to the breakdown of hydrodynamics~\cite{Romatschke:2017ejr}. When the longitudinal pressure of an expanding fluid vanishes during its course of evolution, the fluid breaks into fragments which makes the hydrodynamic description invalid. The condition in which the effective longitudinal pressure of the system becomes negative decides the onset of cavitation. The sharp rise in the bulk viscosity of QGP near the transition temperature $T_c$ can induce cavitation, thereby affecting the fireball evolution and consequently the emitted signals; and this possibility has been studied in several works~\cite{Torrieri:2007fb,Rajagopal:2009yw,Bhatt:2010cy,Bhatt:2010hu,Bhatt:2011kx,Habich:2014tpa,Byres:2019xld,Naik:2022pyk}. Non-occurrence of cavitation scenario was also employed to constrain the value of bulk viscosity in Ref.~\cite{Habich:2014tpa}. 
\par
Shear viscosity of the QGP on the other hand, is expected to have a minimum around $T_c$ and gradually increases with temperature. It was demonstrated by the authors of Ref.~\cite{Bhatt:2011kr} that the presence of shear viscosity can also cause cavitation. 
Using temperature-dependent $\eta/s$ prescriptions, it has been shown that during the initial timescales of fireball evolution, large shear stresses in the medium can lead to negative effective pressures, thus breaking down the hydrodynamic description. 
However, such cavitation scenarios in QGP arising from shear viscosity have not been extensively studied. Importantly, the expanding QGP is known to have robust transverse dynamics and how its presence affects the $\eta/s$ induced cavitation is yet to be investigated. Further, it has been well accepted that the quark-gluon matter formed in the collisions remain highly undersaturated during the early proper times, which would consequently have impact on the evolution, and also on the onset of cavitation, although most of the hydrodynamical studies assume chemically equilibrated matter.
\par
The equilibration of hot QCD matter created in the heavy-ion collisions has been analyzed in several works with great interest~\cite{Matsui:1985eu,Biro:1993qt, Levai:1994dx, Baier:2000sb, Berges:2013eia, Kurkela:2018oqw,Kurkela:2018vqr,Kurkela:2018xxd}. Although the perturbative and kinetic theory treatments signal different timescales for thermalization and chemical equilibration of the matter, both analyses highlight the importance of chemical equilibration dynamics, which subsequently has strong impact on the evolution and observables. The perturbative analysis indicates that the chemical equilibration of the quark-gluon matter occurs after rapid thermalization of the medium, with gluons equilibrating faster owing to the large gluon-gluon cross-sections~\cite{Biro:1993qt,Levai:1994dx,Matsui:1985eu}. Chemical equilibration of the shear viscous QGP has been studied by employing causal relativistic dissipative hydrodynamics~\cite{El:2008yy,Bhatt:2009zg}, and recently, the authors have shown using Gubser geometry that the transverse flow plays a major role in the dynamics~\cite{Naik:2026ehp}. However, the chemical equilibration of radially expanding QGP in presence of temperature-dependent shear viscosity has not been considered before, and we undertake this in the present analysis. 
\par
In this \textit{Letter}, we intend to explore shear viscosity induced cavitation scenarios in a chemically equilibrating QGP using temperature-dependent shear viscosity parameterizations. We model the transverse evolution dynamics of QGP by invoking the Gubser geometry.
After introducing the effect of chemical non-equilibrium in the parton distribution functions through the fugacity parameters, the master rate equations are constructed by considering relevant processes driving the equilibration. The rate equations for these underlying mechanisms are solved simultaneously along with the dynamical evolution equations of the radially expanding viscous plasma. Note that throughout the manuscript, we use $c=\hbar =k_B=1$, and follow the metric signature $(+,-,-,-)$.
\par
We are interested in the hot baryonless quark-gluon matter created in the early stages of heavy-ion collisions. We consider the chemical equilibration of parton gas medium, which has already
reached isotropy in momentum distribution. The chemical non-equilibrium can be represented by the introduction of \textit{fugacities} - $\lambda_i$ into single particle phase-space distributions~\cite{Biro:1993qt}:
\begin{equation}\label{Eq:distribution}
 f_i(p;\lambda_i,T) = \lambda_i \left(e^{\beta\cdot p}\pm \lambda_i \right)^{-1}; 
\end{equation}
where $i \equiv (q, \bar{q}, g)$ and $\beta\cdot p = \beta^\mu p_\mu=T^{-1} u^\mu p_\mu$, with $u^\mu$ being the four-velocity of the comoving frame and $p^\mu$ denoting the four-momenta of the partons. Fugacities lie between $0\leq\lambda_i\leq1$, with $\lambda_i=1$ implying complete chemical equilibration. Since we are interested in the initial stages of the collisions at high energies and the
shear viscosity induced cavitations are known to occur in the early times itself \cite{Bhatt:2011kr}, we work with ultrarelativistic equation of state (EoS) for massless particles {\it i.e.};
$\varepsilon= \varepsilon_q + \varepsilon_{\bar{q}} + \varepsilon_g=3P$ throughout this analysis.
Using the above definition of distribution functions, number ($n$) and energy ($\varepsilon$)
densities of quarks (antiquarks) and gluons can be
calculated as
\begin{eqnarray}
   n_q &=& -\frac{\nu_q T^3}{\pi^2}\textrm{Li}_3[ -\lambda_q] = n_{\bar{q}}, \nonumber \\
   n_g &=& \frac{\nu_g T^3}{\pi^2}\textrm{Li}_3[ +\lambda_g], \nonumber \\
   \varepsilon_q &=& -\frac{3\nu_q T^4}{\pi^2}\textrm{Li}_4[ -\lambda_q] = \varepsilon_{\bar{q}}, \nonumber \\
\varepsilon_g &=& \frac{3\nu_g T^4}{\pi^2}\textrm{Li}_4[ +\lambda_g];\label{Eq:num-en-density}
\end{eqnarray}
where, $\nu_q = 2 N_c N_f$ and $\nu_g = 2(N_c^2 -1)$ are the degeneracy factors of quarks (antiquarks) and gluons, respectively\footnote{The PolyLogarithm of order $n$ and argument $z$ is defined as
\begin{equation*}
    \textrm{Li}_n[z] = \sum_{k = 1}^\infty \frac{z^k}{k^n} \cdot
\end{equation*}}.
We note that in the baryonless QGP that we are considering, quark and antiquark fugacities
are equal: $\lambda_q=\lambda_{\bar{q}}$.
Also, in this work, we take $N_c = 3$ and $N_f = 2.5$.

In the absence of chemical equilibrium, number currents $N^\mu_i=n_i\, u^\mu$ of the different particle species are obtained
from the corresponding \textit{rate equations}
\be
D_\mu N^{\mu}_i\,=R_i;
\ee
where, the rates $R_i$ vanish once the system attains chemical equilibrium.
For the baryonless QGP created in early stages of heavy-ion collisions, the 
prominant reactions driving the equilibration are
$gg \longleftrightarrow ggg$ and $gg \longleftrightarrow q\bar q$ \cite{Matsui:1985eu}. The corresponding
rate equations can be casted as \cite{Biro:1993qt,Bhatt:2011kr}
\begin{eqnarray}
     D_\mu ({n}_q {u}^\mu) &=& {R}_2 {n}_g \left(1- \frac{{n}_q {n}_{\bar{q}} \tilde{n}_g^2}{\tilde{n}_q \tilde{n}_{\bar{q}} {n}_g^2} \right)  \label{Eq:RE1}\\
   {D}_\mu ({n}_g {u}^\mu) &=& {R}_3 {n}_g \left(1- \frac{{n}_g}{\tilde{n}_g} \right) \nonumber \\
   &&- {R}_2{n}_g \left(1- \frac{{n}_q {n}_{\bar{q}} \tilde{{n}}_g^2}{\tilde{{n}}_q \tilde{{n}}_{\bar{q}} {n}_g^2} \right);   \label{Eq:RE2}
\end{eqnarray}
where, $D_\mu$ is the covariant derivative\footnote{
Covariant derivative of a vector $A^\mu$ and a tensor $A^{\mu\nu}$ are defined as
\begin{eqnarray*}
    D_\mu A^\nu &=& \partial_\mu A^\nu + \Gamma_{\mu\sigma}^\nu A^\sigma,  \label{Eq:covariant-der-vector}\\
    D_\alpha A^{\mu\nu} &=& \partial_\alpha A^{\mu\nu} + \Gamma_{\alpha\sigma}^\mu  A^{\nu\sigma} + \Gamma_{\alpha\sigma}^\nu  A^{\mu\sigma};
\end{eqnarray*}
where, Christoffel symbols of second kind are given by,
\begin{eqnarray*}
    \Gamma_{\mu\nu}^\lambda = \frac{g^{\lambda\rho}}{2}
    \left(\partial_\mu g_{\rho\nu} +
    \partial_\nu g_{\rho\mu} -
    \partial_\rho g_{\mu\nu}\right). \label{Eq:christoffel-symbol}
\end{eqnarray*}
}. 
The quantities $\tilde{n}_i$ denote the parton number densities in a fully equilibrated QGP ($\lambda_i=1$). The rates $R_2$ and $R_3$ are the density weighted reaction rates defined in terms of the thermally averaged, velocity weighted cross-sections
$\langle \sigma_{gg\rightarrow ggg} v \rangle$ and $\langle \sigma_{g\rightarrow q} v \rangle$, respectively and are given by~\cite{Biro:1993qt}: $R_2 = \frac{1}{2}\langle \sigma_{gg\rightarrow ggg} v \rangle n_g
\simeq 2.1\alpha^2_sT(2\lambda_g-\lambda^2_g)^{1/2}$
and $R_3= \frac{1}{2}\langle \sigma_{g\rightarrow q} v \rangle n_g
\simeq 0.24N_f\alpha^2_s\lambda_gT\,\ln(1.65/\alpha_s\lambda_g)$; with $\alpha_s$ being the strong coupling constant.
\begin{figure}[t]
 \centering
 \includegraphics[width=\linewidth]{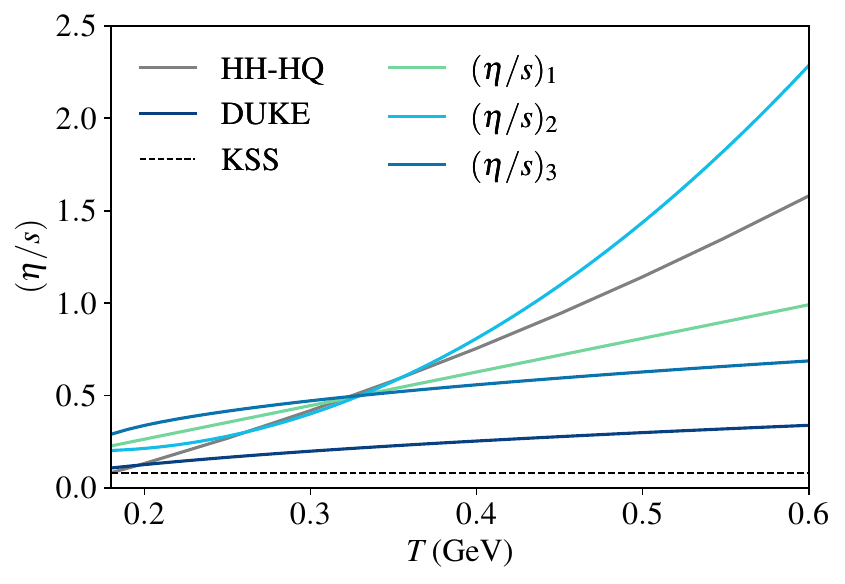}
 \caption{Temperature-dependent shear viscosity prescriptions from Ref.~\cite{Shen:2011eg}, Bayesian analysis (DUKE)~\cite{Bernhard:2019bmu}, high temperature QGP parameterization (HH-HQ) from~\cite{Niemi:2012ry}, along with the KSS bound.}
 \label{fig:etabys}
\end{figure}
\begin{figure*}
 \centering
  \subfigure
  []{\includegraphics[width=0.49\textwidth, height=6.09cm]{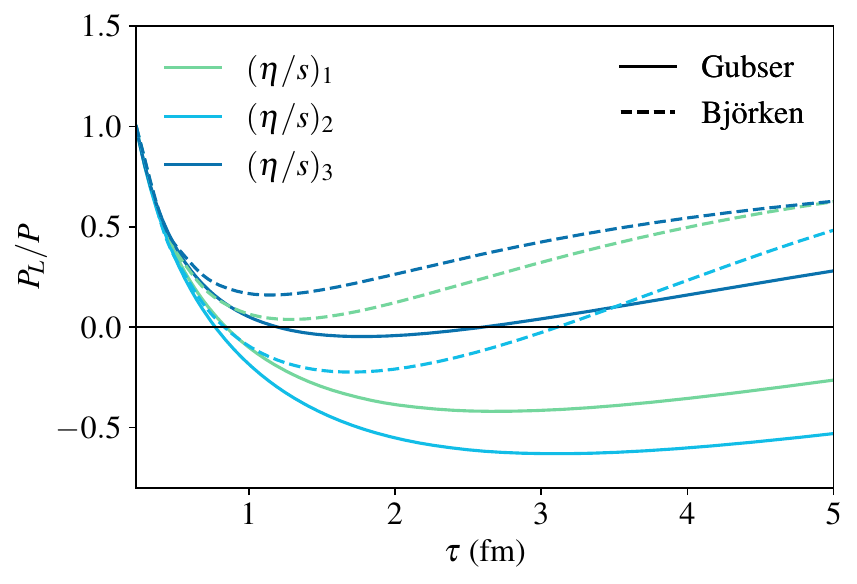}\label{fig:pzbyP}}
  \subfigure[]{\includegraphics[width=0.49\textwidth]{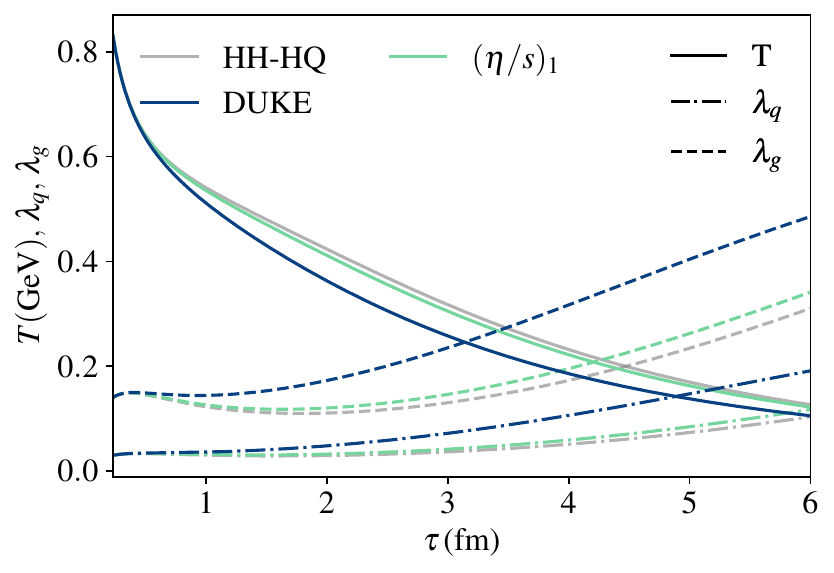}\label{fig:T-lambda}
  }
  \caption{(a) Proper-time evolution of scaled longitudinal pressure obtained from a chemically equilibrating medium within Gubser flow at $r=0$, for different $\eta/s$ parameterizations. The evolution corresponding to the one-dimensional Bj\"orken expansion (dashed curves) are also shown. (b) Evolution of temperature, quark and gluon fugacities of a chemically equilibrating medium within Gubser flow ($r=0$), for different shear viscosity parameterizations. In both the figures, initial conditions are taken to be $T_0 = 0.83$ GeV, $\lambda_q^0 = 0.03$, $\lambda_g^0=0.14$ at $\tau_0=0.23$ fm. }
\end{figure*}
\par
Relativistic hydrodynamical equations in presence of shear viscosity can be obtained from the energy-momentum tensor $T^{\mu\nu}$ as
\ba
D_\mu T^{\mu\nu}&=&0;\,\,\,T^{\mu\nu}=\varepsilon \, u^\mu\,u^\nu\, - P\, \Delta^{\mu\nu} + \pi^{\mu\nu};
\label{Eq:Tmunu}
\ea
where, $\Delta^{\mu\nu} = g^{\mu\nu} - u^\mu u^\nu$ acts as a projection operator perpendicular to $u^\mu$ and $g^{\mu\nu}$ denotes the metric. The tensor $\pi^{\mu\nu}$ represents the shear viscous contribution to $T^{\mu\nu}$ and is symmetric,
traceless ($\pi^\mu_\mu=0$) and orthogonal to the fluid velocity ($u_\mu\pi^{\mu\nu}=0$).
The conservation of $T^{\mu\nu}$ along $u^\mu$ gives the following hydrodynamic equation,
\begin{eqnarray}
D \varepsilon + (\varepsilon+P)\, \Theta-\pi^{\mu\nu}\nabla_{(\mu}\,u_{\nu)}=0;  \label{Eq:edot} 
\end{eqnarray}
where, $D\equiv u^\mu D_\mu$, $\Theta\equiv D_{\mu}\,u^\mu$ is the expansion scalar,
$\nabla_{\alpha}=\Delta_{\mu\alpha} D^{\mu}$
and $A_{(\mu}\,B_{\nu)}=\frac{1}{2}
\left(A_\mu\,B_\nu+A_\nu\,B_\mu\right)$ gives the symmetrization. We use the evolution equation for $\pi^{\mu\nu}$ obtained within the second-order Israel-Stewart formalism. Expanding the entropy four-flux $s^\mu$ up to second order in $\pi^{\mu\nu}$, we get
\begin{eqnarray}
    s^\mu = s u^\mu - \frac{\beta_2}{2T} \pi^{\alpha\beta} \pi_{\alpha\beta} u^\mu.
\end{eqnarray}
Now, employing the second law of thermodynamics gives the evolution for $\pi^{\mu\nu}$~\cite{Muronga:2001zk}
\begin{eqnarray}
 D\pi^{\alpha\beta}
&=& -\frac{1}{\tau_\pi}\pi^{\alpha\beta} +\frac{1}{\beta_2} \nabla^{\langle\alpha} u^{\beta\rangle}
 \nonumber \\
&& - \frac{1}{2\beta_2} \pi^{\alpha\beta} \left(\beta_2 \Theta + TD \left( \frac{\beta_2}{T}\right)\right).\label{Eq:shear-IS}
\end{eqnarray}
Here, the second-order coefficient $\beta_2$ and the shear relaxation time $\tau_{\pi}$ are related
by shear viscosity $\eta$ as $\beta_2= \tau_{\pi}/2\eta$,
and $\nabla_{\langle\alpha} u_{\beta\rangle}\equiv \nabla_
{(\alpha}\,u_{\beta)}-\frac{1}{3}\,\Delta_{\alpha\beta}\Theta$.
Now, the Eqs.~\eqref{Eq:RE1}, \eqref{Eq:RE2}, \eqref{Eq:edot}, and \eqref{Eq:shear-IS} describe the evolution equations for chemically equilibrating
viscous system. 
\par
Several shear viscosity prescriptions for hot QGP formed in
the high energy collisions are used in phenomenological hydrodynamical modeling studies as described earlier.
In this analysis, we use the temperature-dependent shear viscosity prescriptions from Ref.~\cite{Shen:2011eg},
$\left(\eta/s\right)_1 = 0.2 + 0.3\,\frac{T-T_\text{chem}}{T_\text{chem}},\,
\left(\eta/s\right)_2 = 0.2 + 0.4\,\frac{(T-T_\text{chem})^2}{T_\text{chem}^2}\,\text{and}\,
\left(\eta/s\right)_3 = 0.2 + 0.3\,\sqrt{ \frac{T-T_\text{chem}}{T_\text{chem}} }$,
with $T_\text{chem}=0.165$ GeV. Further, 
high temperature QGP viscosity parameterization (HH-HQ) of Ref.~\cite{Niemi:2012ry} and $\eta/s(T)$ based on Bayesian analysis (DUKE)~\cite{Bernhard:2019bmu} are also considered. In Fig.~\ref{fig:etabys}, we plot these shear viscosity prescriptions along with the KSS bound $\eta/s = 1/4\pi$.
\par 
In order to describe the evolution equations within the heavy-ion geometry, we employ Gubser flow which provides semi-realistic radial solutions invoking the conformal invariance $SO(3)$. This model is based on the assumptions that the system possess longitudinal boost invariance and is symmetric under reflections along the spacetime rapdity $\eta_s$ direction. Gubser flow can be described in de Sitter coordinates with Weyl rescaling of the metric using proper-time $\tau$, $ds^2 \rightarrow ds^2/\tau^2 \equiv d\hat{s}^2$, followed by a coordinate transformation to {\it Gubser coordinates} $\rho$ and $\theta$:
\begin{eqnarray}
    \sinh\rho &\equiv& -\frac{1-(q\tau)^2+(qr)^2}{2q\tau}, \label{Eq:rho}\\
    \tan\theta &\equiv& \frac{2qr}{1+(q\tau)^2-(qr)^2};\label{Eq:theta}
\end{eqnarray}
where, $q$ represents the inverse of transverse system size and $ds^2 = d\tau^2 -(dr^2 + r^2 d\phi^2) - \tau^2 d\eta_s^2$ is the metric in Milne coordinates $x^\mu = (\tau, r, \phi, \eta_s)$. The Weyl rescaled line element in the new coordinates $(\rho, \theta, \phi, \eta_s)$ is now given by $ d\hat{s}^2 = d\rho^2 - \cosh^2 \rho\,(d\theta^2 +  \sin^2\theta\,d\phi^2) - d\eta_s^2$. Weyl rescaling renders the fluid homogeneous with all the fields depending only on the coordinate $\rho$. Note that all the quantities depending on the Gubser coordinates are denoted by a {\it hat}. In the de Sitter coordinates, the Gubser flow appears to be static $i.e.$, $\hat{u}^\mu = (1, 0, 0, 0)$. Also, the shear stress tensor can be parameterized as $\hat{\pi}_\nu^\mu = \textrm{diag}(0, \hat{\pi}/2, \hat{\pi}/2, -\hat{\pi})$.
\par
\begin{figure*}[]
    \centering
    \includegraphics[width=\textwidth]{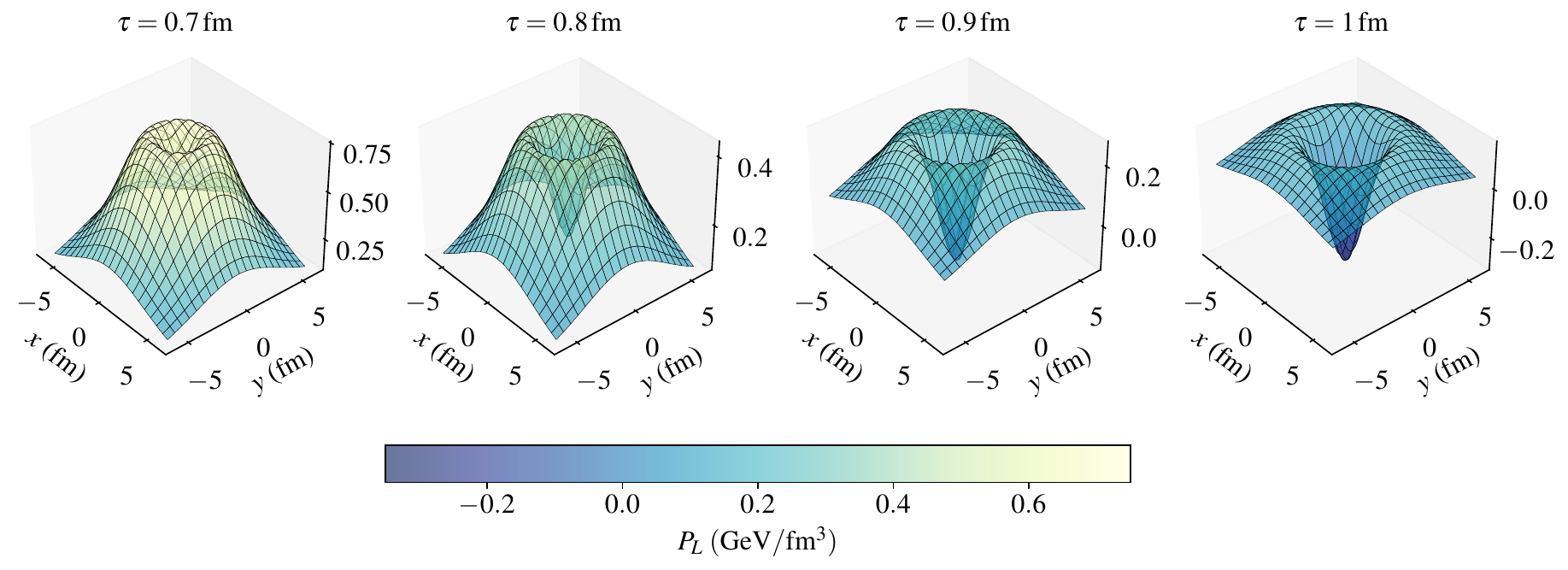}
    \caption{Onset of shear viscosity induced cavitation as a function of radial coordinate in a chemically equilibrating QGP.}
    \label{fig:pz-3d}
\end{figure*}
We cast the evolution equations for energy density (Eq.~\eqref{Eq:edot}), number densities (Eqs.~\eqref{Eq:RE1} and \eqref{Eq:RE2}), and shear pressure (Eq.~\eqref{Eq:shear-IS}) within Gubser coordinates as
\begin{align}
    \frac{4 \That'}{\That} &+ \left(\hat{n}_g \frac{\lambdahat'_g}{\lambdahat_g}   - 2 \hat{n}_q \frac{\lambdahat'_q}{\lambdahat_q}  \right) \frac{\That}{\hat{\varepsilon}} + \left(\frac{4}{3} - \frac{\hat{\pi}}{2\hat{\varepsilon}} \right) \hat{\Theta} = 0,  \nonumber \\
    \hat{\pi}' &+ \frac{\hat{\pi}}{\hat{\tau}_\pi } - \frac{\hat{\Theta}}{3 \hat{\beta}_2} = -\frac{\hat{\pi}}{2 \hat{\beta}_2} \left(\hat{\beta}_2 \hat{\Theta} + \That \frac{d}{d\rho} \left( \frac{\hat{\beta_2}}{\That}\right)  \right), \nonumber  
\end{align}
\begin{align}
     \frac{3 \That'}{\That} &+ \left(\frac{\textrm{Li}_2[-\lambdahat_q]}{\textrm{Li}_3[-\lambdahat_q]} \right) \frac{\lambdahat_q'}{\lambdahat_q} +  \hat{\Theta} \nonumber\\
     &= -\hat{R}_2 \frac{\nu_g}{\nu_q} \left(\frac{\textrm{Li}_3[+\lambdahat_g]}{\textrm{Li}_3[-\lambdahat_q]}\right) \left(1 - \frac{16}{9}\left( \frac{\textrm{Li}_3[-\lambdahat_q]}{\textrm{Li}_3[ +\lambdahat_g]}\right)^2 \right), \nonumber\\
        \frac{3 \That'}{\That} &+ \left(\frac{\textrm{Li}_2[+ \lambdahat_g]}{\textrm{Li}_3[ +\lambdahat_g]}\right) \frac{\lambdahat_g'}{\lambdahat_g} + \hat{\Theta} = \hat{R}_3 \left(1 - \frac{\textrm{Li}_3[ +\lambdahat_g]}{\zeta(3)} \right) \nonumber\\
     &- \hat{R}_2 \left(1 - \frac{16}{9}\left(\frac{\textrm{Li}_3[-\lambdahat_q]}{\textrm{Li}_3[+\lambdahat_g]}\right)^2  \right);
     \label{Eq:evo-gubser}
\end{align}
where, {\it prime} denotes the derivative with respect to the coordinate $\rho$, the expansion scalar is $\hat{\Theta} = 2\tanh\rho$, and $\hat{\pi} = - \hat{\pi}_{\eta_s}^{\eta_s}$. Note that the parton number densities ($\hat{n}_q$ and $\hat{n}_g$) and total energy density ($\hat{\varepsilon}$) are functions of $\rho$ (Eqs.~\eqref{Eq:num-en-density}). These equations are solved simultaneously to obtain the temperature $\That(\rho)$, shear stress $\hat{\pi}(\rho)$ and fugacity $\lambdahat_{q, g}(\rho)$ profiles. To study the cavitation scenarios within the Gubser geometry, we determine the effective pressure of QGP in the longitudinal ($\hat{P}_L$) and transverse ($\hat{P}_T$) directions. From the definition of $\hat{T}_\nu^\mu$ (Eq.~\eqref{Eq:Tmunu}), we get,
\begin{eqnarray}
\hat{P}_L = \hat{P} + \hat{\pi}, \quad\quad
\hat{P}_T = \hat{P} - \frac{1}{2}  \hat{\pi}. \label{Eq:pressures-gubser}
\end{eqnarray}
Here, $\hat{P}(\hat{T},\,\lambdahat_{q, g})$ is the thermodynamic pressure of the system in Gubser coordinates. Note that the effect of shear viscosity is to decrease the longitudinal pressure and increase the transverse one. Now, the cavitation condition is given as $\hat{P}_L = 0$.
The above obtained solutions can be expressed in Milne coordinates through the transformation equations:
\begin{equation}
\begin{aligned}
    T(\tau, r) &= \That(\rho)/\tau,\\
    \lambda_{q, g}(\tau, r) &= \lambdahat_{q, g}(\rho),  \\
  \pi_{\mu\nu}(\tau, r) &= \hat{\pi}_{\alpha\beta} \frac{1}{\tau^2} \frac{\partial \hat{x}^\alpha}{\partial x^\mu} \frac{\partial \hat{x}^\beta}{\partial x^\nu}. \label{Eq:transformation}
\end{aligned}
\end{equation}
\begin{figure*}[t]
    \centering
    \includegraphics[width=\textwidth]{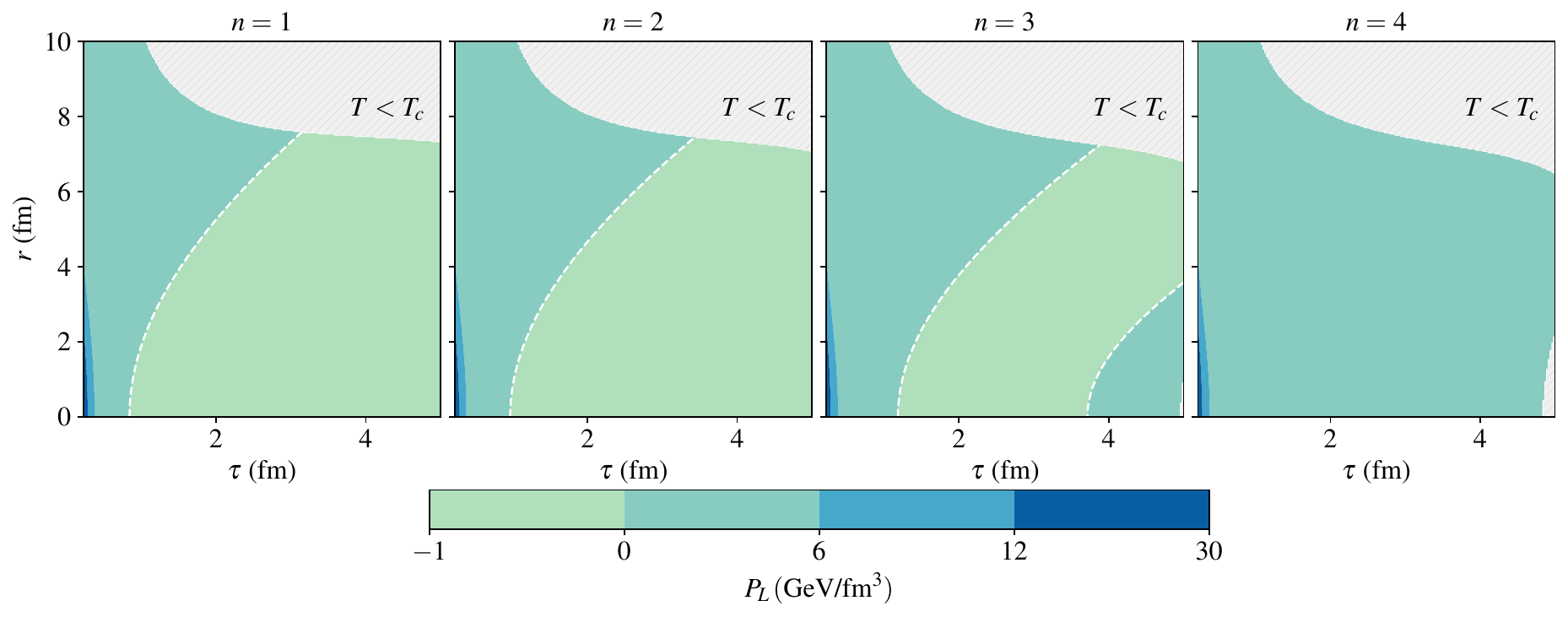}
    \caption{Longitudinal pressure ($P_L$) contours for chemically equilibrating QGP within Gubser flow. Shear viscosity prescription is varied by scaling with $n$: $(\eta/s)_1/n$. The white dashed curves denote the corresponding cavitation boundary.}
       \label{fig:pz-contour-gb}
\end{figure*}
\begin{figure*}
    \centering
    \includegraphics[width=\textwidth]{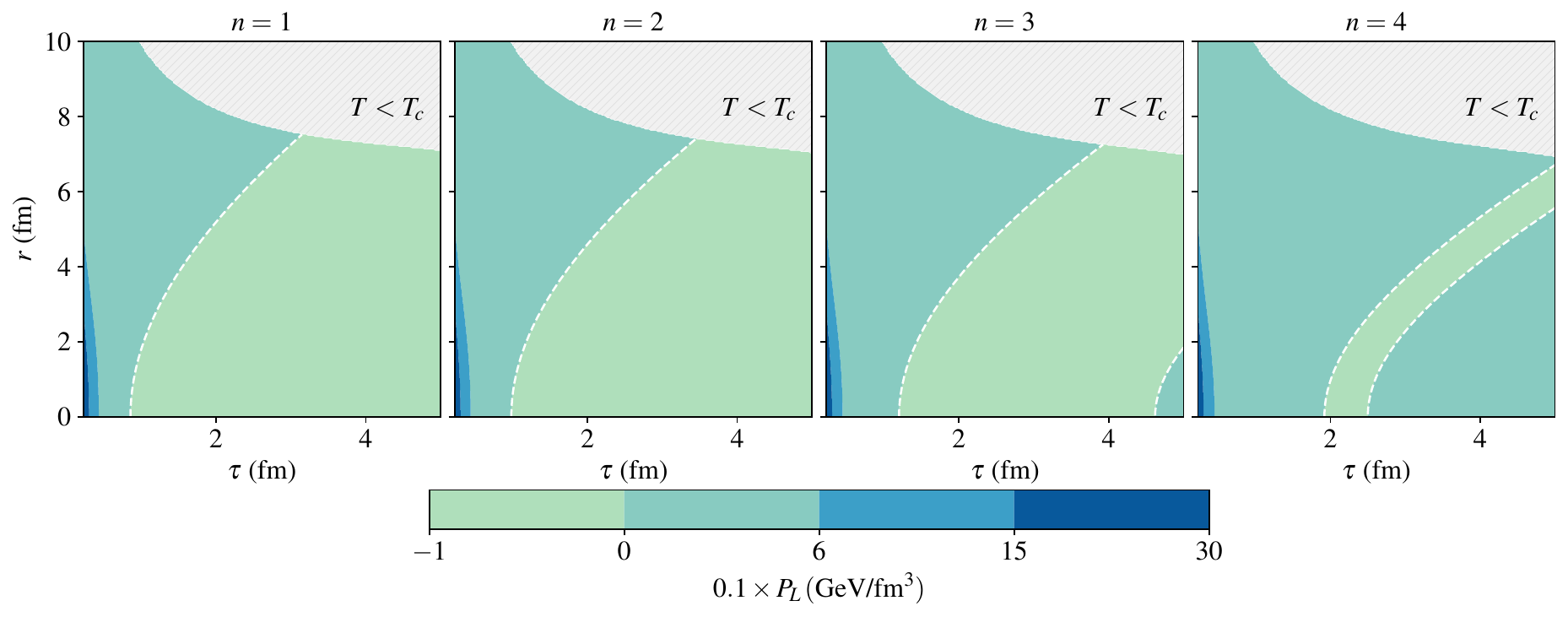}
    \caption{Same as Fig.~\ref{fig:pz-contour-gb}, but for the fully equilibrated ($\lambda_i =1$) medium.}
       \label{fig:pz-contour-gb-eq}
\end{figure*}
\par
We numerically solve the evolution equations described previously by providing relevant initial conditions, and look for the onset of cavitation in the medium. The proper time at which cavitation emerges in the system is represented by $\tau_\text{cav}$. It must be noted that within Gubser flow, the cavitation time depends on the radial coordinate $r$, and this may result in a spatially nonuniform onset of cavitation. Initially, we keep $T_0=0.83$ GeV, $\pi_0 = 0$, $\lambda_q^0=0.03$, and $\lambda_g^0=0.14$ at $(\tau_0, r_0) = ( 0.23$ fm, 0)~\cite{Biro:1993qt, Naik:2026ehp}. The corresponding initial condition in the Gubser coordinates can be determined from Eq.~\eqref{Eq:rho}. We take $q= 1/4.3$ fm$^{-1}$~\cite{Gubser:2010ze} in our analysis and initially the relaxation time is fixed to be $\tau_\pi = 5 (\eta/s)/T$, motivated by the kinetic theory~\cite{Denicol:2010xn}. 
\par
In Fig.~\ref{fig:pzbyP}, we show the evolution of scaled longitudinal pressure ($P_L/P$) in a chemically equilibrating QGP, with transverse flow for the three $\eta/s$ prescriptions from Ref.~\cite{Shen:2011eg}. We compare our results with those obtained in the absence of transverse expansion, employing the one-dimensional longitudinal boost-invariant Bj\"orken flow~\cite{Bjorken:1982qr}. The hydrodynamic evolution equations within the Bj\"orken model have been given in Appendix~\ref{sec:bjorken}. Note that the evolution corresponding to the Gubser flow is plotted for $r=0$. We observe that the chemically equilibrating system with transverse flow undergoes cavitation at very early proper times for all the three $\eta/s$ parameterizations. We find that the inclusion of transverse expansion is crucial for capturing cavitation in the medium, which is absent in the one-dimensional case. 
Further, we observe that the presence of transverse expansion leads to an early onset of cavitation compared to the one-dimensional expansion, as seen from the evolution corresponding to the $(\eta/s)_2$ prescription. 
The cavitation timescales corresponding to different $(\eta/s)$ within Gubser and Bj\"orken models are tabulated in Table~\ref{tab:cavitation}. We also find that varying the relaxation time $\tau_\pi$ does not alter the $\tau_\textrm{cav}$ values substantially, and a larger $\tau_\pi$ brings cavitation at an early timescale, as seen in the presence of bulk viscosity~\cite{Naik:2022pyk}. 

\begin{table}[h]
\centering
\begin{tabular*}{0.5\textwidth}{@{\extracolsep{\fill}} ccccc}
\toprule
\multicolumn{1}{c}{$\eta/s$} & \multicolumn{1}{c}{$\tau_\pi = b (\eta/s)/T$} & \multicolumn{2}{c}{\textbf{Gubser}} & \multicolumn{1}{c}{\textbf{Bj\"orken}} \\
\cmidrule(rl){3-4}
      &      &  $r=0$   & $r = 2$ &    \\
\midrule
\multirow{2}{3em}{$(\eta/s)_1$} & $b = 4$  &   0.88    &  1.06   &  -- \\
                                & $b = 5$  &   0.86    &  1.04   &  -- \\
\midrule
\multirow{2}{3em}{$(\eta/s)_2$} & $b = 4$  &   0.78    &  0.94   &  0.88 \\
                                & $b = 5$  &   0.78    &  0.94   &  0.84 \\
\midrule                                
\multirow{2}{3em}{$(\eta/s)_3$} & $b = 4$  &   --      &  --     &  -- \\
                                & $b = 5$  &   1.22    &  1.45   &  -- \\
\end{tabular*}
\caption{Cavitation time $\tau_{\rm cav}$ (fm) values for different shear viscosity prescriptions plotted in Fig.~\ref{fig:pzbyP} with (Gubser) and without (Bj\"orken) transverse flow.} 
\label{tab:cavitation}
\end{table}

\par
In Fig.~\ref{fig:T-lambda}, we compare the temperature, quark and gluon fugacity evolution profiles obtained for $(\eta/s)_1$, HH-HQ, and DUKE parameterizations. The HH-HQ parameterization brings comparatively large shear viscosity at high temperature. We observe that systems with large value of shear viscosity have slower rate of cooling and lower degree of chemical equilibration. Note that, among the different parameterizations, we use the $(\eta/s)_1$ prescription for the subsequent analysis, as it provides a reasonable $\eta/s$ for the temperature range considered (Fig.~\ref{fig:etabys}). 
\par
We illustrate the onset of cavitation in a chemically equilibrating QGP in presence of transverse expansion by plotting the evolution of longitudinal pressure of the system at different snapshots of proper time in Fig.~\ref{fig:pz-3d}. At early proper times, the longitudinal pressure of the system is positive with maximum value found at the center of the fireball. With increase in $r$, $P_L$ is found to decrease smoothly, since the regions away from the center are cooler. As the system expands, $P_L$ begins to decrease throughout the transverse plane. We find that $P_L$ decreases rapidly around $r=0$ creating a dip in its value around the fireball center. By $\tau = 0.9$ fm, the longitudinal pressure at the center goes to negative, indicating the onset of cavitation in the hotter regions of the fireball, thus giving a fatal blow to the hydrodynamical evolution scheme. Here, cavitation does not occur simultaneously throughout the fireball, but begins at the center and radially permeates to different regimes with increase in proper time. This scenario is in sharp contrast to the bulk viscosity driven case, where cavitation sets in at larger radial distances well before it can advance to the centre of the fireball~\cite{Byres:2019xld}. We tabulate the $\tau_\textrm{cav}$ values at two different radial distances ($r=$ 0, 2 fm) in Table~\ref{tab:cavitation}. 

 \begin{figure}
    \centering
    \includegraphics[width=0.49\textwidth]{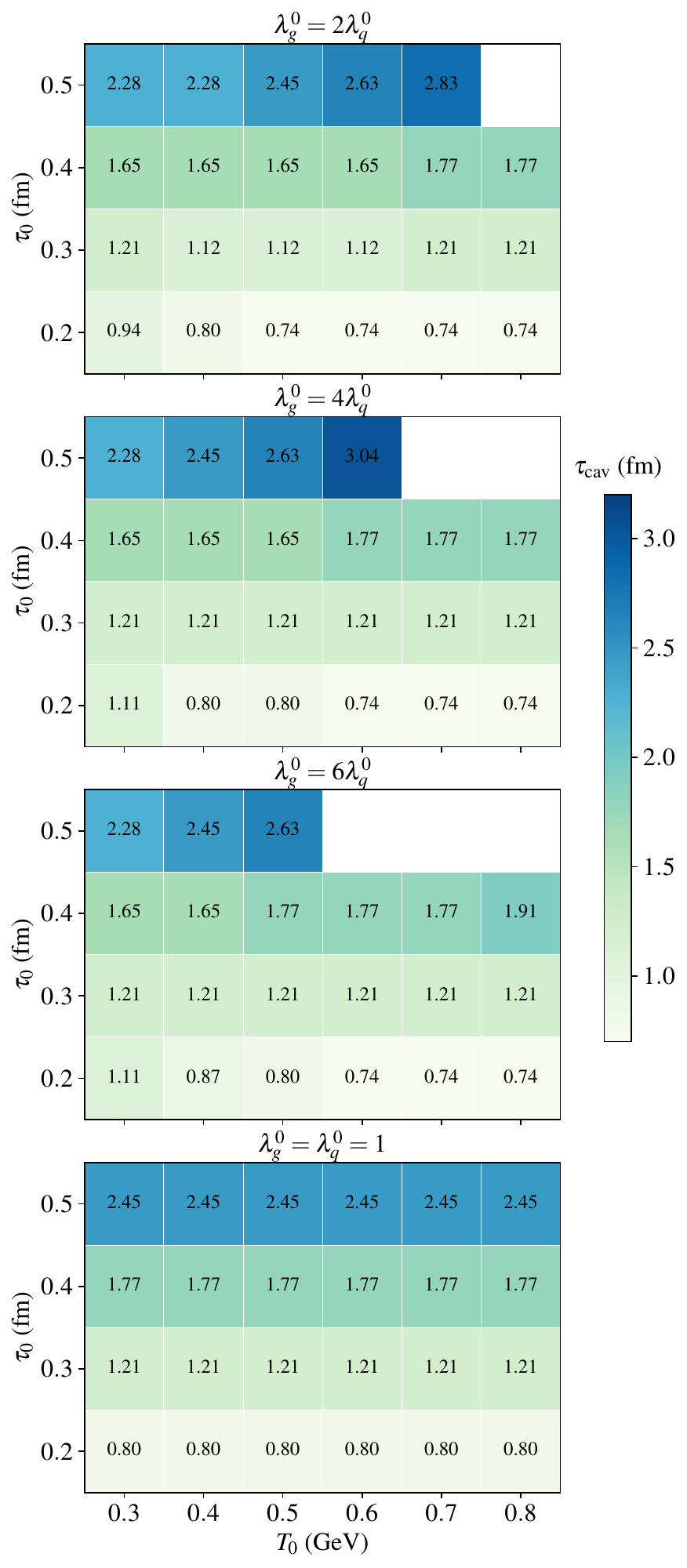}
    \caption{Heat maps representing the cavitation timescales ($\tau_\textrm{cav}$) of the expanding medium by varying the initial temperature and proper time, for different initial composition of partons. The $\tau_\textrm{cav}$ values corresponding to the equilibrated medium ($\lambda_q^0 = \lambda_g^0 = 1$) are also shown. }
    \label{fig:taucav-T-tau}
\end{figure}
\par
In Figs.~\ref{fig:pz-contour-gb} and \ref{fig:pz-contour-gb-eq}, we plot the constant longitudinal pressure contours in the $\tau -r$ plane, obtained from a chemically equilibrating and fully equilibrated QGP respectively. The shear viscosity is scaled as $(\eta/s)_1/n$, and we look for the variation in the cavitation boundary (white dashed curve) with decrease in shear viscosity, by varying $n$. Also, the system is evolved till the critical temperature {\it i.e.,} $T_c=0.15$ GeV. We observe that small value of viscosity (corresponding to large value of $n$) delays the onset of cavitation, and shifts the cavitation boundary to larger $\tau$ values. For the chemically equilibrating system with shear viscosity $(\eta/s)_1/2$, we obtain $\tau_\textrm{cav} = 0.98$ (1.18) fm at $r=0$ (2) fm; while for $(\eta/s)_1/3$, we get $\tau_\textrm{cav} = 1.20$ (1.43) fm at $r=0$ (2) fm. We notice that with further decrease in $\eta/s$, the cavitating region in the $\tau-r$ plane reduces, and below a critical value of $\eta/s$, cavitation is found to disappear from the system. For the initial conditions considered, the value of $(\eta/s)_\textrm{crit}$ is obtained as $(\eta/s)_\textrm{crit} = 0.36\,(0.35)$ for a chemically equilibrating (fully equilibrated) QGP. We also find that the cavitation timescales remain almost equal in both chemical non-equilibrium and equilibrium scenarios. 
\par 
Next, we study the effect of variation of initial conditions on the cavitation timescales in Fig.~\ref{fig:taucav-T-tau}, for a chemically equilibrating medium at $r=0$. The initial gluon fugacity is fixed to be $\lambda_g^0 = 0.14$ and we vary the corresponding values of quark. 
Cavitation scenarios are seen to be robust across different initial conditions considered. We find that, unlike the chemically equilibrated QGP, presence of chemical non-equilibrium tends to remove cavitation from the system for large values of $T_0$ and $\tau_0$. For smaller quark fugacities, the occurrence of cavitation is suppressed in an evolving system. Further, we observe that the cavitation timescales are more sensitive to variations in the initial proper time, compared to the temperature.

\begin{figure}
       \centering
    \includegraphics[width=\linewidth]{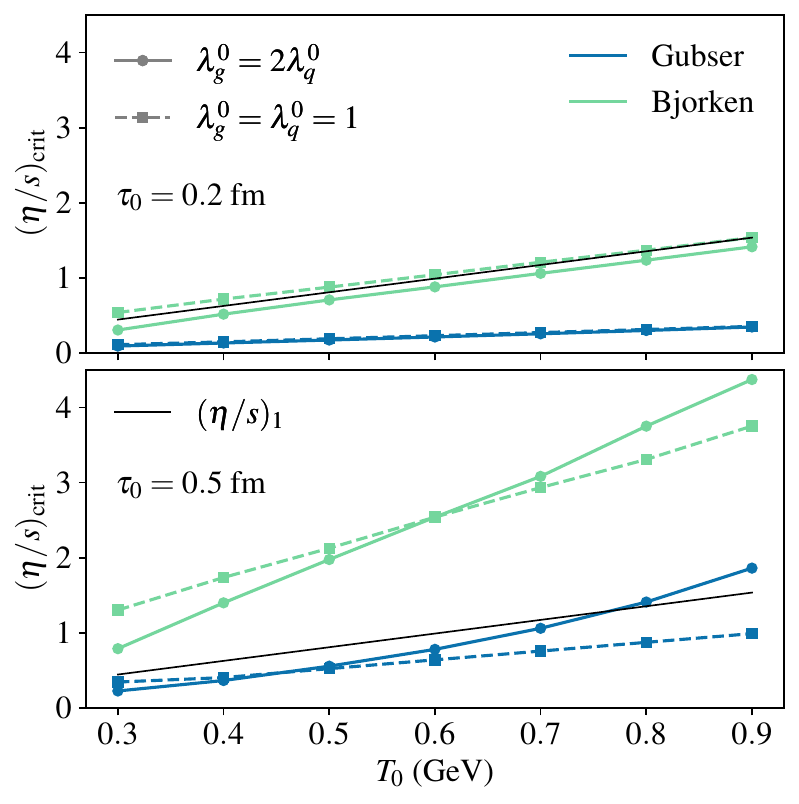}
    \caption{Initial value of $(\eta/s)_\text{crit}$ corresponding to different initial temperatures, varying the initial proper time. The curves corresponding to the equilibrium scenario as well as those obtained within one-dimensional Bj\"orken flow are also shown. The solid black curve represent the $(\eta/s)$ prescription used. }
    \label{fig:etabys-tau0}
\end{figure}

\begin{figure}
       \centering
    \includegraphics[width=\linewidth]{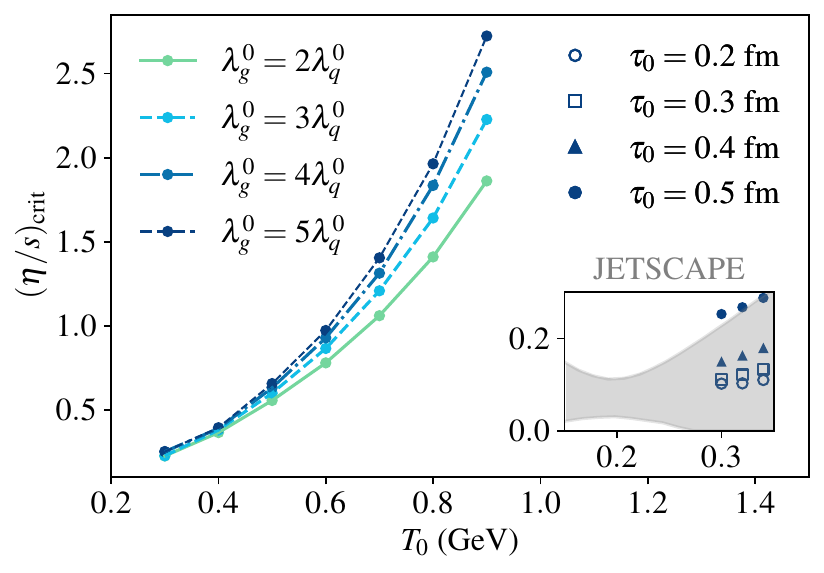}
    \caption{Initial value of ($\eta/s)_\text{crit}$ corresponding to various initial temperatures for $\tau_0=0.5$ fm and varying the initial fugacities. The inset depicts the $(\eta/s)_\text{crit}$ values for different proper time initializations corresponding to $\lambda_g^0 = 5 \lambda_q^0$ case, along with the phenomenological shear viscosity bound from JETSCAPE Bayesian analysis~\cite{JETSCAPE:2020mzn}.} 
    \label{fig:etabys-lg}
\end{figure}

\par
The success of viscous hydrodynamics in describing heavy-ion collision data raises the question of how large the viscous corrections can become before the longitudinal pressure turns negative during the QGP evolution. We therefore determine a critical shear viscosity, $(\eta/s)_{\rm crit}$, above which cavitation occurs. In Fig.~\ref{fig:etabys-tau0}, we show the initial value of $(\eta/s)_\text{crit}$ corresponding to a range of initial temperatures, varying the initial time $\tau_0$ with $\lambda_g^0 = 2 \lambda_q^0$. We also depict the $(\eta/s)_\text{crit}$ values for a longitudinally expanding QGP (green curves) for comparison. The solid black curve represents the shear viscosity prescription $(\eta/s)_1$ used. We find that $(\eta/s)_\text{crit}$ increases with $T_0$, implying that hotter systems can maintain large values of shear viscosity without cavitation. It can also be seen that system initialized at later proper times can sustain large $(\eta/s)$ values. These results are in line with that observed for the case of bulk viscous cavitation in Ref.~\cite{Habich:2014tpa}. Further, QGP initialized at early times supports small values of viscosity under chemical non-equilibrium compared to the equilibrium case. Whereas, the system can sustain higher values of $\eta/s$ at large $T_0$ compared to the equilibrium scenario when initialized at late proper times. We also find that the presence of transverse expansion suppress the allowed values of shear viscosity the system can maintain without cavitation. 
\par 
Finally, in Fig.~\ref{fig:etabys-lg}, we plot the initial $(\eta/s)_\text{crit}$ values obtained for a non-equilibrium QGP undergoing transverse expansion, by varying the initial value of quark fugacities. We fix $\tau_0=0.5$ fm and $\lambda_g^0=0.14$ for this analysis. It can be seen that the system initialised with a smaller quark density can sustain larger viscosity without cavitating, especially at higher initial temperatures. In the inset of Fig.~\ref{fig:etabys-lg}, we plot the $(\eta/s)_\text{crit}$ varying the initial proper time for $\lambda_g^0=5\lambda_q^0$. Also shown is the phenomenological $(\eta/s)$ bound from JETSCAPE Bayesian analysis for the Chapman-Enskog model~\cite{JETSCAPE:2020mzn}. We see that the $(\eta/s)_\text{crit}$ values corresponding to larger $\tau_0$ lie above the bound, indicating that the phenomenologically extracted values remain below the critical threshold and therefore do not lead to cavitation within the present framework. 
Our analyses demonstrates that $(\eta/s)_\text{crit}$ values are strongly dependent on the initial conditions chosen. The interplay between chemical non-equilibrium, transverse expansion, and cooling of the medium plays a major role in constraining the allowed values of shear viscosity of the system. 
\par
In summary, we have studied the shear viscosity induced cavitation at the early times in a radially expanding QGP by taking into the account the effect of chemical undersaturation. The effect of chemical non-equilibrium is introduced in the particle distribution functions through fugacities, and the master rate equations are solved along with the causal dissipative relativistic hydrodynamic evolution equations by considering different phenomenological temperature-dependent shear viscosity parameterizations. We observe that large values of shear viscosity drive the system to negative pressure scenarios at the early times of expansion itself. We illustrated that shear induced cavitation first develops at the centre of the fireball and subsequently extends to larger transverse distances, in contrast to the behavior observed for bulk viscosity. The transverse flow dynamics is found to accelerate the cavitation. 
We found that the onset of cavitation is sensitive to the initial temperature and proper time as well as the initial composition of partons. Our results indicate that shear induced cavitation persist amidst chemical equilibration dynamics of the system. We note that tracking the onset of $P_L < 0$ is necessary in successful hydrodynamic modeling of the shear viscous QGP evolution.

Further, within the considered framework, we constrained the allowed values of the shear viscosity by requiring the absence of cavitation during the QGP evolution. We found that the presence of transverse flow suppresses the maximum allowed $\eta/s$ values of the medium. We also demonstrated that systems initialised at larger proper times with strong quark undersaturation can sustain large initial shear viscosity. Note that the present analysis employs a conformal EoS and the symmetry based Gubser geometry. Consequently, the obtained critical initial $\eta/s$ values should be interpreted as model-dependent thresholds rather than quantitative bounds on the QGP transport coefficient. A realistic EoS and full numerical hydrodynamic evolution of the QGP would be required for a direct phenomenological constraint.

\appendix

\section{Hydrodynamic evolution within Bj\"orken flow} \label{sec:bjorken}

In the Milne coordinates $(\tau, x, y, \eta_s)$, the shear stress tensor is parameterized as $\pi_\nu^{\mu} = \textrm{diag}(0, -\pi/2, -\pi/2, \pi)$. Noting the fluid four-velocity within Bj\"orken geometry to be $u^\mu = (1, 0, 0, 0)$, we cast the $T(\tau), \pi(\tau), \lambda_q(\tau)$, $\lambda_g(\tau)$ evolution equations for chemically equilibrating, longitudinally expanding viscous QGP as
\begin{align}
\frac{4\dot{T}}{T} &+
\left(n_g \frac{\dot{\lambda}_g}{\lambda_g} - 2 n_q \frac{\dot{\lambda}_q}{\lambda_q} \right) \frac{T}{\varepsilon} + \left(\frac{4}{3} - \frac{\pi}{\varepsilon} \right) \Theta = 0, \nonumber\\
\dot{\pi} &+ \frac{\pi}{\tau_\pi} - \frac{2}{3}\frac{\Theta}{\beta_2} = - \frac{\pi}{2\beta_2} \left(\beta_2 \Theta + T \frac{d}{d\tau} \left(\frac{\beta_2}{T} \right) \right), \nonumber  \\
3\frac{\dot{T}}{T} &+ \left(\frac{\textrm{Li}_2[ +\lambda_q]}{\textrm{Li}_3[ +\lambda_q]}\right)   \frac{\dot{\lambda_q}}{\lambda_q} + \Theta  \nonumber \\
&= -R_2 \frac{\nu_g}{\nu_q} \left(\frac{\textrm{Li}_3[ +\lambda_g]}{\textrm{Li}_3[ -\lambda_q]}\right) \left(1 - \frac{16}{9}\left(\frac{\textrm{Li}_3[ -\lambda_q]}{\textrm{Li}_3[ +\lambda_g]} \right)^2 \right),
  \nonumber  \\
3\frac{\dot{T}}{T} &+ \left(\frac{\textrm{Li}_2[ +\lambda_g]}{\textrm{Li}_3[ +\lambda_g]}\right)  \frac{\dot{\lambda_g}}{\lambda_g}+ \Theta
= R_3 \left(1 - \frac{\textrm{Li}_3[ +\lambda_g]}{\zeta(3)} \right) \nonumber \\
     &- R_2 \left(1 - \frac{16}{9}\left(\frac{\textrm{Li}_3[ -\lambda_q]}{\textrm{Li}_3[+ \lambda_g]}\right)^2 \right),
 \label{Eq:evo-bjorken}
\end{align}
respectively. Here, the proper time derivative is denoted by the {\it dot} and $\Theta = 1/\tau$. Within Bj\"orken flow, the transverse and longitudinal pressures of the expanding fluid are respectively given by
\begin{equation}
P_L = P - \pi, \quad \quad
P_T = P + \frac{1}{2} \pi ; \label{Eq:pressures-bjorken}
\end{equation}
with $P_L=0$ being the condition for cavitation.

\bibliography{Cav-ChemEq}

\begin{thebibliography}{63}%
\makeatletter
\providecommand \@ifxundefined [1]{%
 \@ifx{#1\undefined}
}%
\providecommand \@ifnum [1]{%
 \ifnum #1\expandafter \@firstoftwo
 \else \expandafter \@secondoftwo
 \fi
}%
\providecommand \@ifx [1]{%
 \ifx #1\expandafter \@firstoftwo
 \else \expandafter \@secondoftwo
 \fi
}%
\providecommand \natexlab [1]{#1}%
\providecommand \enquote  [1]{``#1''}%
\providecommand \bibnamefont  [1]{#1}%
\providecommand \bibfnamefont [1]{#1}%
\providecommand \citenamefont [1]{#1}%
\providecommand \href@noop [0]{\@secondoftwo}%
\providecommand \href [0]{\begingroup \@sanitize@url \@href}%
\providecommand \@href[1]{\@@startlink{#1}\@@href}%
\providecommand \@@href[1]{\endgroup#1\@@endlink}%
\providecommand \@sanitize@url [0]{\catcode `\\12\catcode `\$12\catcode
  `\&12\catcode `\#12\catcode `\^12\catcode `\_12\catcode `\%12\relax}%
\providecommand \@@startlink[1]{}%
\providecommand \@@endlink[0]{}%
\providecommand \url  [0]{\begingroup\@sanitize@url \@url }%
\providecommand \@url [1]{\endgroup\@href {#1}{\urlprefix }}%
\providecommand \urlprefix  [0]{URL }%
\providecommand \Eprint [0]{\href }%
\providecommand \doibase [0]{https://doi.org/}%
\providecommand \selectlanguage [0]{\@gobble}%
\providecommand \bibinfo  [0]{\@secondoftwo}%
\providecommand \bibfield  [0]{\@secondoftwo}%
\providecommand \translation [1]{[#1]}%
\providecommand \BibitemOpen [0]{}%
\providecommand \bibitemStop [0]{}%
\providecommand \bibitemNoStop [0]{.\EOS\space}%
\providecommand \EOS [0]{\spacefactor3000\relax}%
\providecommand \BibitemShut  [1]{\csname bibitem#1\endcsname}%
\let\auto@bib@innerbib\@empty
\bibitem [{\citenamefont {Meyer}(2011)}]{Meyer:2011gj}%
  \BibitemOpen
  \bibfield  {author} {\bibinfo {author} {\bibfnamefont {H.~B.}\ \bibnamefont
  {Meyer}},\ }\bibfield  {title} {\bibinfo {title} {{Transport Properties of
  the Quark-Gluon Plasma: A Lattice QCD Perspective}},\ }\href
  {https://doi.org/10.1140/epja/i2011-11086-3} {\bibfield  {journal} {\bibinfo
  {journal} {Eur. Phys. J. A}\ }\textbf {\bibinfo {volume} {47}},\ \bibinfo
  {pages} {86} (\bibinfo {year} {2011})},\ \Eprint
  {https://arxiv.org/abs/1104.3708} {arXiv:1104.3708 [hep-lat]} \BibitemShut
  {NoStop}%
\bibitem [{\citenamefont {Braun-Munzinger}\ \emph {et~al.}(2016)\citenamefont
  {Braun-Munzinger}, \citenamefont {Koch}, \citenamefont {Sch{\"a}fer},\ and\
  \citenamefont {Stachel}}]{Braun-Munzinger:2015hba}%
  \BibitemOpen
  \bibfield  {author} {\bibinfo {author} {\bibfnamefont {P.}~\bibnamefont
  {Braun-Munzinger}}, \bibinfo {author} {\bibfnamefont {V.}~\bibnamefont
  {Koch}}, \bibinfo {author} {\bibfnamefont {T.}~\bibnamefont {Sch{\"a}fer}},\
  and\ \bibinfo {author} {\bibfnamefont {J.}~\bibnamefont {Stachel}},\
  }\bibfield  {title} {\bibinfo {title} {{Properties of hot and dense matter
  from relativistic heavy ion collisions}},\ }\href
  {https://doi.org/10.1016/j.physrep.2015.12.003} {\bibfield  {journal}
  {\bibinfo  {journal} {Phys. Rept.}\ }\textbf {\bibinfo {volume} {621}},\
  \bibinfo {pages} {76} (\bibinfo {year} {2016})},\ \Eprint
  {https://arxiv.org/abs/1510.00442} {arXiv:1510.00442 [nucl-th]} \BibitemShut
  {NoStop}%
\bibitem [{\citenamefont {Shuryak}(2017)}]{Shuryak:2014zxa}%
  \BibitemOpen
  \bibfield  {author} {\bibinfo {author} {\bibfnamefont {E.}~\bibnamefont
  {Shuryak}},\ }\bibfield  {title} {\bibinfo {title} {{Strongly coupled
  quark-gluon plasma in heavy ion collisions}},\ }\href
  {https://doi.org/10.1103/RevModPhys.89.035001} {\bibfield  {journal}
  {\bibinfo  {journal} {Rev. Mod. Phys.}\ }\textbf {\bibinfo {volume} {89}},\
  \bibinfo {pages} {035001} (\bibinfo {year} {2017})},\ \Eprint
  {https://arxiv.org/abs/1412.8393} {arXiv:1412.8393 [hep-ph]} \BibitemShut
  {NoStop}%
\bibitem [{\citenamefont {Jaiswal}\ \emph {et~al.}(2021)\citenamefont {Jaiswal}
  \emph {et~al.}}]{Jaiswal:2020hvk}%
  \BibitemOpen
  \bibfield  {author} {\bibinfo {author} {\bibfnamefont {A.}~\bibnamefont
  {Jaiswal}} \emph {et~al.},\ }\bibfield  {title} {\bibinfo {title} {{Dynamics
  of QCD matter {\textemdash} current status}},\ }\href
  {https://doi.org/10.1142/S0218301321300010} {\bibfield  {journal} {\bibinfo
  {journal} {Int. J. Mod. Phys. E}\ }\textbf {\bibinfo {volume} {30}},\
  \bibinfo {pages} {2130001} (\bibinfo {year} {2021})},\ \Eprint
  {https://arxiv.org/abs/2007.14959} {arXiv:2007.14959 [hep-ph]} \BibitemShut
  {NoStop}%
\bibitem [{\citenamefont {Hirano}\ and\ \citenamefont
  {Gyulassy}(2006)}]{Hirano:2005wx}%
  \BibitemOpen
  \bibfield  {author} {\bibinfo {author} {\bibfnamefont {T.}~\bibnamefont
  {Hirano}}\ and\ \bibinfo {author} {\bibfnamefont {M.}~\bibnamefont
  {Gyulassy}},\ }\bibfield  {title} {\bibinfo {title} {{Perfect fluidity of the
  quark gluon plasma core as seen through its dissipative hadronic corona}},\
  }\href {https://doi.org/10.1016/j.nuclphysa.2006.02.005} {\bibfield
  {journal} {\bibinfo  {journal} {Nucl. Phys. A}\ }\textbf {\bibinfo {volume}
  {769}},\ \bibinfo {pages} {71} (\bibinfo {year} {2006})},\ \Eprint
  {https://arxiv.org/abs/nucl-th/0506049} {arXiv:nucl-th/0506049} \BibitemShut
  {NoStop}%
\bibitem [{\citenamefont {Romatschke}\ and\ \citenamefont
  {Romatschke}(2007)}]{Romatschke:2007mq}%
  \BibitemOpen
  \bibfield  {author} {\bibinfo {author} {\bibfnamefont {P.}~\bibnamefont
  {Romatschke}}\ and\ \bibinfo {author} {\bibfnamefont {U.}~\bibnamefont
  {Romatschke}},\ }\bibfield  {title} {\bibinfo {title} {{Viscosity information
  from relativistic nuclear collisions: how perfect is the fluid observed at
  RHIC?}},\ }\href {https://doi.org/10.1103/PhysRevLett.99.172301} {\bibfield
  {journal} {\bibinfo  {journal} {Phys. Rev. Lett.}\ }\textbf {\bibinfo
  {volume} {99}},\ \bibinfo {pages} {172301} (\bibinfo {year} {2007})},\
  \Eprint {https://arxiv.org/abs/0706.1522} {arXiv:0706.1522 [nucl-th]}
  \BibitemShut {NoStop}%
\bibitem [{\citenamefont {Heinz}\ and\ \citenamefont
  {Snellings}(2013)}]{Heinz:2013th}%
  \BibitemOpen
  \bibfield  {author} {\bibinfo {author} {\bibfnamefont {U.}~\bibnamefont
  {Heinz}}\ and\ \bibinfo {author} {\bibfnamefont {R.}~\bibnamefont
  {Snellings}},\ }\bibfield  {title} {\bibinfo {title} {{Collective flow and
  viscosity in relativistic heavy-ion collisions}},\ }\href
  {https://doi.org/10.1146/annurev-nucl-102212-170540} {\bibfield  {journal}
  {\bibinfo  {journal} {Ann. Rev. Nucl. Part. Sci.}\ }\textbf {\bibinfo
  {volume} {63}},\ \bibinfo {pages} {123} (\bibinfo {year} {2013})},\ \Eprint
  {https://arxiv.org/abs/1301.2826} {arXiv:1301.2826 [nucl-th]} \BibitemShut
  {NoStop}%
\bibitem [{\citenamefont {Kovtun}\ \emph {et~al.}(2005)\citenamefont {Kovtun},
  \citenamefont {Son},\ and\ \citenamefont {Starinets}}]{Kovtun:2004de}%
  \BibitemOpen
  \bibfield  {author} {\bibinfo {author} {\bibfnamefont {P.}~\bibnamefont
  {Kovtun}}, \bibinfo {author} {\bibfnamefont {D.~T.}\ \bibnamefont {Son}},\
  and\ \bibinfo {author} {\bibfnamefont {A.~O.}\ \bibnamefont {Starinets}},\
  }\bibfield  {title} {\bibinfo {title} {{Viscosity in strongly interacting
  quantum field theories from black hole physics}},\ }\href
  {https://doi.org/10.1103/PhysRevLett.94.111601} {\bibfield  {journal}
  {\bibinfo  {journal} {Phys. Rev. Lett.}\ }\textbf {\bibinfo {volume} {94}},\
  \bibinfo {pages} {111601} (\bibinfo {year} {2005})},\ \Eprint
  {https://arxiv.org/abs/hep-th/0405231} {arXiv:hep-th/0405231} \BibitemShut
  {NoStop}%
\bibitem [{\citenamefont {Arnold}\ \emph {et~al.}(2003)\citenamefont {Arnold},
  \citenamefont {Moore},\ and\ \citenamefont {Yaffe}}]{Arnold:2003zc}%
  \BibitemOpen
  \bibfield  {author} {\bibinfo {author} {\bibfnamefont {P.~B.}\ \bibnamefont
  {Arnold}}, \bibinfo {author} {\bibfnamefont {G.~D.}\ \bibnamefont {Moore}},\
  and\ \bibinfo {author} {\bibfnamefont {L.~G.}\ \bibnamefont {Yaffe}},\
  }\bibfield  {title} {\bibinfo {title} {{Transport coefficients in high
  temperature gauge theories. 2. Beyond leading log}},\ }\href
  {https://doi.org/10.1088/1126-6708/2003/05/051} {\bibfield  {journal}
  {\bibinfo  {journal} {JHEP}\ }\textbf {\bibinfo {volume} {05}},\ \bibinfo
  {pages} {051}},\ \Eprint {https://arxiv.org/abs/hep-ph/0302165}
  {arXiv:hep-ph/0302165} \BibitemShut {NoStop}%
\bibitem [{\citenamefont {Csernai}\ \emph {et~al.}(2006)\citenamefont
  {Csernai}, \citenamefont {Kapusta},\ and\ \citenamefont
  {McLerran}}]{Csernai:2006zz}%
  \BibitemOpen
  \bibfield  {author} {\bibinfo {author} {\bibfnamefont {L.~P.}\ \bibnamefont
  {Csernai}}, \bibinfo {author} {\bibfnamefont {J.~I.}\ \bibnamefont
  {Kapusta}},\ and\ \bibinfo {author} {\bibfnamefont {L.~D.}\ \bibnamefont
  {McLerran}},\ }\bibfield  {title} {\bibinfo {title} {{On the
  Strongly-Interacting Low-Viscosity Matter Created in Relativistic Nuclear
  Collisions}},\ }\href {https://doi.org/10.1103/PhysRevLett.97.152303}
  {\bibfield  {journal} {\bibinfo  {journal} {Phys. Rev. Lett.}\ }\textbf
  {\bibinfo {volume} {97}},\ \bibinfo {pages} {152303} (\bibinfo {year}
  {2006})},\ \Eprint {https://arxiv.org/abs/nucl-th/0604032}
  {arXiv:nucl-th/0604032} \BibitemShut {NoStop}%
\bibitem [{\citenamefont {Lacey}\ \emph {et~al.}(2007)\citenamefont {Lacey},
  \citenamefont {Ajitanand}, \citenamefont {Alexander}, \citenamefont {Chung},
  \citenamefont {Holzmann}, \citenamefont {Issah}, \citenamefont {Taranenko},
  \citenamefont {Danielewicz},\ and\ \citenamefont {Stoecker}}]{Lacey:2006bc}%
  \BibitemOpen
  \bibfield  {author} {\bibinfo {author} {\bibfnamefont {R.~A.}\ \bibnamefont
  {Lacey}}, \bibinfo {author} {\bibfnamefont {N.~N.}\ \bibnamefont
  {Ajitanand}}, \bibinfo {author} {\bibfnamefont {J.~M.}\ \bibnamefont
  {Alexander}}, \bibinfo {author} {\bibfnamefont {P.}~\bibnamefont {Chung}},
  \bibinfo {author} {\bibfnamefont {W.~G.}\ \bibnamefont {Holzmann}}, \bibinfo
  {author} {\bibfnamefont {M.}~\bibnamefont {Issah}}, \bibinfo {author}
  {\bibfnamefont {A.}~\bibnamefont {Taranenko}}, \bibinfo {author}
  {\bibfnamefont {P.}~\bibnamefont {Danielewicz}},\ and\ \bibinfo {author}
  {\bibfnamefont {H.}~\bibnamefont {Stoecker}},\ }\bibfield  {title} {\bibinfo
  {title} {{Has the QCD Critical Point been Signaled by Observations at
  RHIC?}},\ }\href {https://doi.org/10.1103/PhysRevLett.98.092301} {\bibfield
  {journal} {\bibinfo  {journal} {Phys. Rev. Lett.}\ }\textbf {\bibinfo
  {volume} {98}},\ \bibinfo {pages} {092301} (\bibinfo {year} {2007})},\
  \Eprint {https://arxiv.org/abs/nucl-ex/0609025} {arXiv:nucl-ex/0609025}
  \BibitemShut {NoStop}%
\bibitem [{\citenamefont {Niemi}\ \emph {et~al.}(2011)\citenamefont {Niemi},
  \citenamefont {Denicol}, \citenamefont {Huovinen}, \citenamefont {Molnar},\
  and\ \citenamefont {Rischke}}]{Niemi:2011ix}%
  \BibitemOpen
  \bibfield  {author} {\bibinfo {author} {\bibfnamefont {H.}~\bibnamefont
  {Niemi}}, \bibinfo {author} {\bibfnamefont {G.~S.}\ \bibnamefont {Denicol}},
  \bibinfo {author} {\bibfnamefont {P.}~\bibnamefont {Huovinen}}, \bibinfo
  {author} {\bibfnamefont {E.}~\bibnamefont {Molnar}},\ and\ \bibinfo {author}
  {\bibfnamefont {D.~H.}\ \bibnamefont {Rischke}},\ }\bibfield  {title}
  {\bibinfo {title} {{Influence of the shear viscosity of the quark-gluon
  plasma on elliptic flow in ultrarelativistic heavy-ion collisions}},\ }\href
  {https://doi.org/10.1103/PhysRevLett.106.212302} {\bibfield  {journal}
  {\bibinfo  {journal} {Phys. Rev. Lett.}\ }\textbf {\bibinfo {volume} {106}},\
  \bibinfo {pages} {212302} (\bibinfo {year} {2011})},\ \Eprint
  {https://arxiv.org/abs/1101.2442} {arXiv:1101.2442 [nucl-th]} \BibitemShut
  {NoStop}%
\bibitem [{\citenamefont {Denicol}\ \emph {et~al.}(2016)\citenamefont
  {Denicol}, \citenamefont {Monnai},\ and\ \citenamefont
  {Schenke}}]{Denicol:2015nhu}%
  \BibitemOpen
  \bibfield  {author} {\bibinfo {author} {\bibfnamefont {G.}~\bibnamefont
  {Denicol}}, \bibinfo {author} {\bibfnamefont {A.}~\bibnamefont {Monnai}},\
  and\ \bibinfo {author} {\bibfnamefont {B.}~\bibnamefont {Schenke}},\
  }\bibfield  {title} {\bibinfo {title} {{Moving forward to constrain the shear
  viscosity of QCD matter}},\ }\href
  {https://doi.org/10.1103/PhysRevLett.116.212301} {\bibfield  {journal}
  {\bibinfo  {journal} {Phys. Rev. Lett.}\ }\textbf {\bibinfo {volume} {116}},\
  \bibinfo {pages} {212301} (\bibinfo {year} {2016})},\ \Eprint
  {https://arxiv.org/abs/1512.01538} {arXiv:1512.01538 [nucl-th]} \BibitemShut
  {NoStop}%
\bibitem [{\citenamefont {Vujanovic}\ \emph {et~al.}(2018)\citenamefont
  {Vujanovic}, \citenamefont {Denicol}, \citenamefont {Luzum}, \citenamefont
  {Jeon},\ and\ \citenamefont {Gale}}]{Vujanovic:2017psb}%
  \BibitemOpen
  \bibfield  {author} {\bibinfo {author} {\bibfnamefont {G.}~\bibnamefont
  {Vujanovic}}, \bibinfo {author} {\bibfnamefont {G.~S.}\ \bibnamefont
  {Denicol}}, \bibinfo {author} {\bibfnamefont {M.}~\bibnamefont {Luzum}},
  \bibinfo {author} {\bibfnamefont {S.}~\bibnamefont {Jeon}},\ and\ \bibinfo
  {author} {\bibfnamefont {C.}~\bibnamefont {Gale}},\ }\bibfield  {title}
  {\bibinfo {title} {{Investigating the temperature dependence of the specific
  shear viscosity of QCD matter with dilepton radiation}},\ }\href
  {https://doi.org/10.1103/PhysRevC.98.014902} {\bibfield  {journal} {\bibinfo
  {journal} {Phys. Rev. C}\ }\textbf {\bibinfo {volume} {98}},\ \bibinfo
  {pages} {014902} (\bibinfo {year} {2018})},\ \Eprint
  {https://arxiv.org/abs/1702.02941} {arXiv:1702.02941 [nucl-th]} \BibitemShut
  {NoStop}%
\bibitem [{\citenamefont {Meyer}(2007)}]{Meyer:2007ic}%
  \BibitemOpen
  \bibfield  {author} {\bibinfo {author} {\bibfnamefont {H.~B.}\ \bibnamefont
  {Meyer}},\ }\bibfield  {title} {\bibinfo {title} {{A Calculation of the shear
  viscosity in SU(3) gluodynamics}},\ }\href
  {https://doi.org/10.1103/PhysRevD.76.101701} {\bibfield  {journal} {\bibinfo
  {journal} {Phys. Rev. D}\ }\textbf {\bibinfo {volume} {76}},\ \bibinfo
  {pages} {101701} (\bibinfo {year} {2007})},\ \Eprint
  {https://arxiv.org/abs/0704.1801} {arXiv:0704.1801 [hep-lat]} \BibitemShut
  {NoStop}%
\bibitem [{\citenamefont {Fuini}\ \emph {et~al.}(2011)\citenamefont {Fuini},
  \citenamefont {Demir}, \citenamefont {Srivastava},\ and\ \citenamefont
  {Bass}}]{Fuini:2010xz}%
  \BibitemOpen
  \bibfield  {author} {\bibinfo {author} {\bibfnamefont {J.}~\bibnamefont
  {Fuini}, \bibfnamefont {III}}, \bibinfo {author} {\bibfnamefont {N.~S.}\
  \bibnamefont {Demir}}, \bibinfo {author} {\bibfnamefont {D.~K.}\ \bibnamefont
  {Srivastava}},\ and\ \bibinfo {author} {\bibfnamefont {S.~A.}\ \bibnamefont
  {Bass}},\ }\bibfield  {title} {\bibinfo {title} {{Shear Viscosity in a
  Perturbative Quark-Gluon-Plasma}},\ }\href
  {https://doi.org/10.1088/0954-3899/38/1/015004} {\bibfield  {journal}
  {\bibinfo  {journal} {J. Phys. G}\ }\textbf {\bibinfo {volume} {38}},\
  \bibinfo {pages} {015004} (\bibinfo {year} {2011})},\ \Eprint
  {https://arxiv.org/abs/1008.2306} {arXiv:1008.2306 [nucl-th]} \BibitemShut
  {NoStop}%
\bibitem [{\citenamefont {Chen}\ \emph {et~al.}(2010)\citenamefont {Chen},
  \citenamefont {Dong}, \citenamefont {Ohnishi},\ and\ \citenamefont
  {Wang}}]{Chen:2009sm}%
  \BibitemOpen
  \bibfield  {author} {\bibinfo {author} {\bibfnamefont {J.-W.}\ \bibnamefont
  {Chen}}, \bibinfo {author} {\bibfnamefont {H.}~\bibnamefont {Dong}}, \bibinfo
  {author} {\bibfnamefont {K.}~\bibnamefont {Ohnishi}},\ and\ \bibinfo {author}
  {\bibfnamefont {Q.}~\bibnamefont {Wang}},\ }\bibfield  {title} {\bibinfo
  {title} {{Shear Viscosity of a Gluon Plasma in Perturbative QCD}},\ }\href
  {https://doi.org/10.1016/j.physletb.2010.01.072} {\bibfield  {journal}
  {\bibinfo  {journal} {Phys. Lett. B}\ }\textbf {\bibinfo {volume} {685}},\
  \bibinfo {pages} {277} (\bibinfo {year} {2010})},\ \Eprint
  {https://arxiv.org/abs/0907.2486} {arXiv:0907.2486 [nucl-th]} \BibitemShut
  {NoStop}%
\bibitem [{\citenamefont {Christiansen}\ \emph {et~al.}(2015)\citenamefont
  {Christiansen}, \citenamefont {Haas}, \citenamefont {Pawlowski},\ and\
  \citenamefont {Strodthoff}}]{Christiansen:2014ypa}%
  \BibitemOpen
  \bibfield  {author} {\bibinfo {author} {\bibfnamefont {N.}~\bibnamefont
  {Christiansen}}, \bibinfo {author} {\bibfnamefont {M.}~\bibnamefont {Haas}},
  \bibinfo {author} {\bibfnamefont {J.~M.}\ \bibnamefont {Pawlowski}},\ and\
  \bibinfo {author} {\bibfnamefont {N.}~\bibnamefont {Strodthoff}},\ }\bibfield
   {title} {\bibinfo {title} {{Transport Coefficients in Yang--Mills Theory and
  QCD}},\ }\href {https://doi.org/10.1103/PhysRevLett.115.112002} {\bibfield
  {journal} {\bibinfo  {journal} {Phys. Rev. Lett.}\ }\textbf {\bibinfo
  {volume} {115}},\ \bibinfo {pages} {112002} (\bibinfo {year} {2015})},\
  \Eprint {https://arxiv.org/abs/1411.7986} {arXiv:1411.7986 [hep-ph]}
  \BibitemShut {NoStop}%
\bibitem [{\citenamefont {Astrakhantsev}\ \emph {et~al.}(2017)\citenamefont
  {Astrakhantsev}, \citenamefont {Braguta},\ and\ \citenamefont
  {Kotov}}]{Astrakhantsev:2017nrs}%
  \BibitemOpen
  \bibfield  {author} {\bibinfo {author} {\bibfnamefont {N.}~\bibnamefont
  {Astrakhantsev}}, \bibinfo {author} {\bibfnamefont {V.}~\bibnamefont
  {Braguta}},\ and\ \bibinfo {author} {\bibfnamefont {A.}~\bibnamefont
  {Kotov}},\ }\bibfield  {title} {\bibinfo {title} {{Temperature dependence of
  shear viscosity of $SU(3)$--gluodynamics within lattice simulation}},\ }\href
  {https://doi.org/10.1007/JHEP04(2017)101} {\bibfield  {journal} {\bibinfo
  {journal} {JHEP}\ }\textbf {\bibinfo {volume} {04}},\ \bibinfo {pages}
  {101}},\ \Eprint {https://arxiv.org/abs/1701.02266} {arXiv:1701.02266
  [hep-lat]} \BibitemShut {NoStop}%
\bibitem [{\citenamefont {Bors{\'a}nyi}\ \emph {et~al.}(2018)\citenamefont
  {Bors{\'a}nyi}, \citenamefont {Fodor}, \citenamefont {Giordano},
  \citenamefont {Katz}, \citenamefont {Pasztor}, \citenamefont {Ratti},
  \citenamefont {Sch{\"a}fer}, \citenamefont {Szabo},\ and\ \citenamefont
  {T{\'o}th}}]{Borsanyi:2018srz}%
  \BibitemOpen
  \bibfield  {author} {\bibinfo {author} {\bibfnamefont {S.}~\bibnamefont
  {Bors{\'a}nyi}}, \bibinfo {author} {\bibfnamefont {Z.}~\bibnamefont {Fodor}},
  \bibinfo {author} {\bibfnamefont {M.}~\bibnamefont {Giordano}}, \bibinfo
  {author} {\bibfnamefont {S.~D.}\ \bibnamefont {Katz}}, \bibinfo {author}
  {\bibfnamefont {A.}~\bibnamefont {Pasztor}}, \bibinfo {author} {\bibfnamefont
  {C.}~\bibnamefont {Ratti}}, \bibinfo {author} {\bibfnamefont
  {A.}~\bibnamefont {Sch{\"a}fer}}, \bibinfo {author} {\bibfnamefont {K.~K.}\
  \bibnamefont {Szabo}},\ and\ \bibinfo {author} {\bibfnamefont
  {B.}~\bibnamefont {T{\'o}th}},\ }\bibfield  {title} {\bibinfo {title} {{High
  statistics lattice study of stress tensor correlators in pure $SU(3)$ gauge
  theory}},\ }\href {https://doi.org/10.1103/PhysRevD.98.014512} {\bibfield
  {journal} {\bibinfo  {journal} {Phys. Rev. D}\ }\textbf {\bibinfo {volume}
  {98}},\ \bibinfo {pages} {014512} (\bibinfo {year} {2018})},\ \Eprint
  {https://arxiv.org/abs/1802.07718} {arXiv:1802.07718 [hep-lat]} \BibitemShut
  {NoStop}%
\bibitem [{\citenamefont {Liu}\ and\ \citenamefont {Rapp}(2020)}]{Liu:2016ysz}%
  \BibitemOpen
  \bibfield  {author} {\bibinfo {author} {\bibfnamefont {S.~Y.~F.}\
  \bibnamefont {Liu}}\ and\ \bibinfo {author} {\bibfnamefont {R.}~\bibnamefont
  {Rapp}},\ }\bibfield  {title} {\bibinfo {title} {{Spectral and transport
  properties of quark{\textendash}gluon plasma in a nonperturbative
  approach}},\ }\href {https://doi.org/10.1140/epja/s10050-020-00024-z}
  {\bibfield  {journal} {\bibinfo  {journal} {Eur. Phys. J. A}\ }\textbf
  {\bibinfo {volume} {56}},\ \bibinfo {pages} {44} (\bibinfo {year} {2020})},\
  \Eprint {https://arxiv.org/abs/1612.09138} {arXiv:1612.09138 [nucl-th]}
  \BibitemShut {NoStop}%
\bibitem [{\citenamefont {Ohanaka}\ and\ \citenamefont
  {Lin}(2026)}]{Ohanaka:2026hjx}%
  \BibitemOpen
  \bibfield  {author} {\bibinfo {author} {\bibfnamefont {O.}~\bibnamefont
  {Ohanaka}}\ and\ \bibinfo {author} {\bibfnamefont {Z.-W.}\ \bibnamefont
  {Lin}},\ }\bibfield  {title} {\bibinfo {title} {{Shear viscosity of a
  massless quark-gluon gas in chemical equilibrium in terms of all
  2{\ensuremath{\leftrightarrow}}2 cross sections}},\ }\href
  {https://doi.org/10.1103/878b-dyqp} {\bibfield  {journal} {\bibinfo
  {journal} {Phys. Rev. D}\ }\textbf {\bibinfo {volume} {114}},\ \bibinfo
  {pages} {034019} (\bibinfo {year} {2026})},\ \Eprint
  {https://arxiv.org/abs/2602.08155} {arXiv:2602.08155 [hep-ph]} \BibitemShut
  {NoStop}%
\bibitem [{\citenamefont {Shen}\ \emph {et~al.}(2011)\citenamefont {Shen},
  \citenamefont {Heinz}, \citenamefont {Huovinen},\ and\ \citenamefont
  {Song}}]{Shen:2011eg}%
  \BibitemOpen
  \bibfield  {author} {\bibinfo {author} {\bibfnamefont {C.}~\bibnamefont
  {Shen}}, \bibinfo {author} {\bibfnamefont {U.}~\bibnamefont {Heinz}},
  \bibinfo {author} {\bibfnamefont {P.}~\bibnamefont {Huovinen}},\ and\
  \bibinfo {author} {\bibfnamefont {H.}~\bibnamefont {Song}},\ }\bibfield
  {title} {\bibinfo {title} {{Radial and elliptic flow in Pb+Pb collisions at
  the Large Hadron Collider from viscous hydrodynamic}},\ }\href
  {https://doi.org/10.1103/PhysRevC.84.044903} {\bibfield  {journal} {\bibinfo
  {journal} {Phys. Rev. C}\ }\textbf {\bibinfo {volume} {84}},\ \bibinfo
  {pages} {044903} (\bibinfo {year} {2011})},\ \Eprint
  {https://arxiv.org/abs/1105.3226} {arXiv:1105.3226 [nucl-th]} \BibitemShut
  {NoStop}%
\bibitem [{\citenamefont {Niemi}\ \emph {et~al.}(2012)\citenamefont {Niemi},
  \citenamefont {Denicol}, \citenamefont {Huovinen}, \citenamefont {Molnar},\
  and\ \citenamefont {Rischke}}]{Niemi:2012ry}%
  \BibitemOpen
  \bibfield  {author} {\bibinfo {author} {\bibfnamefont {H.}~\bibnamefont
  {Niemi}}, \bibinfo {author} {\bibfnamefont {G.~S.}\ \bibnamefont {Denicol}},
  \bibinfo {author} {\bibfnamefont {P.}~\bibnamefont {Huovinen}}, \bibinfo
  {author} {\bibfnamefont {E.}~\bibnamefont {Molnar}},\ and\ \bibinfo {author}
  {\bibfnamefont {D.~H.}\ \bibnamefont {Rischke}},\ }\bibfield  {title}
  {\bibinfo {title} {{Influence of a temperature-dependent shear viscosity on
  the azimuthal asymmetries of transverse momentum spectra in ultrarelativistic
  heavy-ion collisions}},\ }\href {https://doi.org/10.1103/PhysRevC.86.014909}
  {\bibfield  {journal} {\bibinfo  {journal} {Phys. Rev. C}\ }\textbf {\bibinfo
  {volume} {86}},\ \bibinfo {pages} {014909} (\bibinfo {year} {2012})},\
  \Eprint {https://arxiv.org/abs/1203.2452} {arXiv:1203.2452 [nucl-th]}
  \BibitemShut {NoStop}%
\bibitem [{\citenamefont {Bernhard}\ \emph {et~al.}(2016)\citenamefont
  {Bernhard}, \citenamefont {Moreland}, \citenamefont {Bass}, \citenamefont
  {Liu},\ and\ \citenamefont {Heinz}}]{Bernhard:2016tnd}%
  \BibitemOpen
  \bibfield  {author} {\bibinfo {author} {\bibfnamefont {J.~E.}\ \bibnamefont
  {Bernhard}}, \bibinfo {author} {\bibfnamefont {J.~S.}\ \bibnamefont
  {Moreland}}, \bibinfo {author} {\bibfnamefont {S.~A.}\ \bibnamefont {Bass}},
  \bibinfo {author} {\bibfnamefont {J.}~\bibnamefont {Liu}},\ and\ \bibinfo
  {author} {\bibfnamefont {U.}~\bibnamefont {Heinz}},\ }\bibfield  {title}
  {\bibinfo {title} {{Applying Bayesian parameter estimation to relativistic
  heavy-ion collisions: simultaneous characterization of the initial state and
  quark-gluon plasma medium}},\ }\href
  {https://doi.org/10.1103/PhysRevC.94.024907} {\bibfield  {journal} {\bibinfo
  {journal} {Phys. Rev. C}\ }\textbf {\bibinfo {volume} {94}},\ \bibinfo
  {pages} {024907} (\bibinfo {year} {2016})},\ \Eprint
  {https://arxiv.org/abs/1605.03954} {arXiv:1605.03954 [nucl-th]} \BibitemShut
  {NoStop}%
\bibitem [{\citenamefont {Bernhard}\ \emph {et~al.}(2019)\citenamefont
  {Bernhard}, \citenamefont {Moreland},\ and\ \citenamefont
  {Bass}}]{Bernhard:2019bmu}%
  \BibitemOpen
  \bibfield  {author} {\bibinfo {author} {\bibfnamefont {J.~E.}\ \bibnamefont
  {Bernhard}}, \bibinfo {author} {\bibfnamefont {J.~S.}\ \bibnamefont
  {Moreland}},\ and\ \bibinfo {author} {\bibfnamefont {S.~A.}\ \bibnamefont
  {Bass}},\ }\bibfield  {title} {\bibinfo {title} {{Bayesian estimation of the
  specific shear and bulk viscosity of quark{\textendash}gluon plasma}},\
  }\href {https://doi.org/10.1038/s41567-019-0611-8} {\bibfield  {journal}
  {\bibinfo  {journal} {Nature Phys.}\ }\textbf {\bibinfo {volume} {15}},\
  \bibinfo {pages} {1113} (\bibinfo {year} {2019})}\BibitemShut {NoStop}%
\bibitem [{\citenamefont {Parkkila}\ \emph {et~al.}(2021)\citenamefont
  {Parkkila}, \citenamefont {Onnerstad},\ and\ \citenamefont
  {Kim}}]{Parkkila:2021tqq}%
  \BibitemOpen
  \bibfield  {author} {\bibinfo {author} {\bibfnamefont {J.~E.}\ \bibnamefont
  {Parkkila}}, \bibinfo {author} {\bibfnamefont {A.}~\bibnamefont
  {Onnerstad}},\ and\ \bibinfo {author} {\bibfnamefont {D.~J.}\ \bibnamefont
  {Kim}},\ }\bibfield  {title} {\bibinfo {title} {{Bayesian estimation of the
  specific shear and bulk viscosity of the quark-gluon plasma with additional
  flow harmonic observables}},\ }\href
  {https://doi.org/10.1103/PhysRevC.104.054904} {\bibfield  {journal} {\bibinfo
   {journal} {Phys. Rev. C}\ }\textbf {\bibinfo {volume} {104}},\ \bibinfo
  {pages} {054904} (\bibinfo {year} {2021})},\ \Eprint
  {https://arxiv.org/abs/2106.05019} {arXiv:2106.05019 [hep-ph]} \BibitemShut
  {NoStop}%
\bibitem [{\citenamefont {Everett}\ \emph {et~al.}(2021)\citenamefont {Everett}
  \emph {et~al.}}]{JETSCAPE:2020mzn}%
  \BibitemOpen
  \bibfield  {author} {\bibinfo {author} {\bibfnamefont {D.}~\bibnamefont
  {Everett}} \emph {et~al.} (\bibinfo {collaboration} {JETSCAPE}),\ }\bibfield
  {title} {\bibinfo {title} {{Multisystem Bayesian constraints on the transport
  coefficients of QCD matter}},\ }\href
  {https://doi.org/10.1103/PhysRevC.103.054904} {\bibfield  {journal} {\bibinfo
   {journal} {Phys. Rev. C}\ }\textbf {\bibinfo {volume} {103}},\ \bibinfo
  {pages} {054904} (\bibinfo {year} {2021})},\ \Eprint
  {https://arxiv.org/abs/2011.01430} {arXiv:2011.01430 [hep-ph]} \BibitemShut
  {NoStop}%
\bibitem [{\citenamefont {Nijs}\ \emph {et~al.}(2021)\citenamefont {Nijs},
  \citenamefont {van~der Schee}, \citenamefont {G{\"u}rsoy},\ and\
  \citenamefont {Snellings}}]{Nijs:2020ors}%
  \BibitemOpen
  \bibfield  {author} {\bibinfo {author} {\bibfnamefont {G.}~\bibnamefont
  {Nijs}}, \bibinfo {author} {\bibfnamefont {W.}~\bibnamefont {van~der Schee}},
  \bibinfo {author} {\bibfnamefont {U.}~\bibnamefont {G{\"u}rsoy}},\ and\
  \bibinfo {author} {\bibfnamefont {R.}~\bibnamefont {Snellings}},\ }\bibfield
  {title} {\bibinfo {title} {{Transverse Momentum Differential Global Analysis
  of Heavy-Ion Collisions}},\ }\href
  {https://doi.org/10.1103/PhysRevLett.126.202301} {\bibfield  {journal}
  {\bibinfo  {journal} {Phys. Rev. Lett.}\ }\textbf {\bibinfo {volume} {126}},\
  \bibinfo {pages} {202301} (\bibinfo {year} {2021})},\ \Eprint
  {https://arxiv.org/abs/2010.15130} {arXiv:2010.15130 [nucl-th]} \BibitemShut
  {NoStop}%
\bibitem [{\citenamefont {Parkkila}\ \emph {et~al.}(2022)\citenamefont
  {Parkkila}, \citenamefont {Onnerstad}, \citenamefont {Taghavi}, \citenamefont
  {Mordasini}, \citenamefont {Bilandzic}, \citenamefont {Virta},\ and\
  \citenamefont {Kim}}]{Parkkila:2021yha}%
  \BibitemOpen
  \bibfield  {author} {\bibinfo {author} {\bibfnamefont {J.~E.}\ \bibnamefont
  {Parkkila}}, \bibinfo {author} {\bibfnamefont {A.}~\bibnamefont {Onnerstad}},
  \bibinfo {author} {\bibfnamefont {S.~F.}\ \bibnamefont {Taghavi}}, \bibinfo
  {author} {\bibfnamefont {C.}~\bibnamefont {Mordasini}}, \bibinfo {author}
  {\bibfnamefont {A.}~\bibnamefont {Bilandzic}}, \bibinfo {author}
  {\bibfnamefont {M.}~\bibnamefont {Virta}},\ and\ \bibinfo {author}
  {\bibfnamefont {D.~J.}\ \bibnamefont {Kim}},\ }\bibfield  {title} {\bibinfo
  {title} {{New constraints for QCD matter from improved Bayesian parameter
  estimation in heavy-ion collisions at LHC}},\ }\href
  {https://doi.org/10.1016/j.physletb.2022.137485} {\bibfield  {journal}
  {\bibinfo  {journal} {Phys. Lett. B}\ }\textbf {\bibinfo {volume} {835}},\
  \bibinfo {pages} {137485} (\bibinfo {year} {2022})},\ \Eprint
  {https://arxiv.org/abs/2111.08145} {arXiv:2111.08145 [hep-ph]} \BibitemShut
  {NoStop}%
\bibitem [{\citenamefont {Heffernan}\ \emph {et~al.}(2024)\citenamefont
  {Heffernan}, \citenamefont {Gale}, \citenamefont {Jeon},\ and\ \citenamefont
  {Paquet}}]{Heffernan:2023utr}%
  \BibitemOpen
  \bibfield  {author} {\bibinfo {author} {\bibfnamefont {M.~R.}\ \bibnamefont
  {Heffernan}}, \bibinfo {author} {\bibfnamefont {C.}~\bibnamefont {Gale}},
  \bibinfo {author} {\bibfnamefont {S.}~\bibnamefont {Jeon}},\ and\ \bibinfo
  {author} {\bibfnamefont {J.-F.}\ \bibnamefont {Paquet}},\ }\bibfield  {title}
  {\bibinfo {title} {{Bayesian quantification of strongly interacting matter
  with color glass condensate initial conditions}},\ }\href
  {https://doi.org/10.1103/PhysRevC.109.065207} {\bibfield  {journal} {\bibinfo
   {journal} {Phys. Rev. C}\ }\textbf {\bibinfo {volume} {109}},\ \bibinfo
  {pages} {065207} (\bibinfo {year} {2024})},\ \Eprint
  {https://arxiv.org/abs/2302.09478} {arXiv:2302.09478 [nucl-th]} \BibitemShut
  {NoStop}%
\bibitem [{\citenamefont {Luzum}\ and\ \citenamefont
  {Romatschke}(2008)}]{Luzum:2008cw}%
  \BibitemOpen
  \bibfield  {author} {\bibinfo {author} {\bibfnamefont {M.}~\bibnamefont
  {Luzum}}\ and\ \bibinfo {author} {\bibfnamefont {P.}~\bibnamefont
  {Romatschke}},\ }\bibfield  {title} {\bibinfo {title} {{Conformal
  Relativistic Viscous Hydrodynamics: Applications to RHIC results at
  s(NN)**(1/2) = 200-GeV}},\ }\href
  {https://doi.org/10.1103/PhysRevC.78.034915} {\bibfield  {journal} {\bibinfo
  {journal} {Phys. Rev. C}\ }\textbf {\bibinfo {volume} {78}},\ \bibinfo
  {pages} {034915} (\bibinfo {year} {2008})},\ \bibinfo {note} {[Erratum:
  Phys.Rev.C 79, 039903 (2009)]},\ \Eprint {https://arxiv.org/abs/0804.4015}
  {arXiv:0804.4015 [nucl-th]} \BibitemShut {NoStop}%
\bibitem [{\citenamefont {Romatschke}\ and\ \citenamefont
  {Romatschke}(2019)}]{Romatschke:2017ejr}%
  \BibitemOpen
  \bibfield  {author} {\bibinfo {author} {\bibfnamefont {P.}~\bibnamefont
  {Romatschke}}\ and\ \bibinfo {author} {\bibfnamefont {U.}~\bibnamefont
  {Romatschke}},\ }\href {https://doi.org/10.1017/9781108651998} {\emph
  {\bibinfo {title} {{Relativistic Fluid Dynamics In and Out of
  Equilibrium}}}},\ Cambridge Monographs on Mathematical Physics\ (\bibinfo
  {publisher} {Cambridge University Press},\ \bibinfo {year} {2019})\ \Eprint
  {https://arxiv.org/abs/1712.05815} {arXiv:1712.05815 [nucl-th]} \BibitemShut
  {NoStop}%
\bibitem [{\citenamefont {Busza}\ \emph {et~al.}(2018)\citenamefont {Busza},
  \citenamefont {Rajagopal},\ and\ \citenamefont {van~der
  Schee}}]{Busza:2018rrf}%
  \BibitemOpen
  \bibfield  {author} {\bibinfo {author} {\bibfnamefont {W.}~\bibnamefont
  {Busza}}, \bibinfo {author} {\bibfnamefont {K.}~\bibnamefont {Rajagopal}},\
  and\ \bibinfo {author} {\bibfnamefont {W.}~\bibnamefont {van~der Schee}},\
  }\bibfield  {title} {\bibinfo {title} {{Heavy Ion Collisions: The Big
  Picture, and the Big Questions}},\ }\href
  {https://doi.org/10.1146/annurev-nucl-101917-020852} {\bibfield  {journal}
  {\bibinfo  {journal} {Ann. Rev. Nucl. Part. Sci.}\ }\textbf {\bibinfo
  {volume} {68}},\ \bibinfo {pages} {339} (\bibinfo {year} {2018})},\ \Eprint
  {https://arxiv.org/abs/1802.04801} {arXiv:1802.04801 [hep-ph]} \BibitemShut
  {NoStop}%
\bibitem [{\citenamefont {Landau}\ and\ \citenamefont
  {Lifshitz}(1987)}]{Landau:1987}%
  \BibitemOpen
  \bibfield  {author} {\bibinfo {author} {\bibfnamefont {L.~D.}\ \bibnamefont
  {Landau}}\ and\ \bibinfo {author} {\bibfnamefont {E.~M.}\ \bibnamefont
  {Lifshitz}},\ }\href@noop {} {\emph {\bibinfo {title} {{Fluid Mechanics}}}}\
  (\bibinfo {year} {1987})\BibitemShut {NoStop}%
\bibitem [{\citenamefont {Eckart}(1940)}]{Eckart:1940te}%
  \BibitemOpen
  \bibfield  {author} {\bibinfo {author} {\bibfnamefont {C.}~\bibnamefont
  {Eckart}},\ }\bibfield  {title} {\bibinfo {title} {{The Thermodynamics of
  irreversible processes. 3.. Relativistic theory of the simple fluid}},\
  }\href {https://doi.org/10.1103/PhysRev.58.919} {\bibfield  {journal}
  {\bibinfo  {journal} {Phys. Rev.}\ }\textbf {\bibinfo {volume} {58}},\
  \bibinfo {pages} {919} (\bibinfo {year} {1940})}\BibitemShut {NoStop}%
\bibitem [{\citenamefont {Hiscock}\ and\ \citenamefont
  {Lindblom}(1983)}]{Hiscock:1983zz}%
  \BibitemOpen
  \bibfield  {author} {\bibinfo {author} {\bibfnamefont {W.~A.}\ \bibnamefont
  {Hiscock}}\ and\ \bibinfo {author} {\bibfnamefont {L.}~\bibnamefont
  {Lindblom}},\ }\bibfield  {title} {\bibinfo {title} {{Stability and causality
  in dissipative relativistic fluids}},\ }\href
  {https://doi.org/10.1016/0003-4916(83)90288-9} {\bibfield  {journal}
  {\bibinfo  {journal} {Annals Phys.}\ }\textbf {\bibinfo {volume} {151}},\
  \bibinfo {pages} {466} (\bibinfo {year} {1983})}\BibitemShut {NoStop}%
\bibitem [{\citenamefont {Hiscock}\ and\ \citenamefont
  {Lindblom}(1985)}]{Hiscock:1985zz}%
  \BibitemOpen
  \bibfield  {author} {\bibinfo {author} {\bibfnamefont {W.~A.}\ \bibnamefont
  {Hiscock}}\ and\ \bibinfo {author} {\bibfnamefont {L.}~\bibnamefont
  {Lindblom}},\ }\bibfield  {title} {\bibinfo {title} {{Generic instabilities
  in first-order dissipative relativistic fluid theories}},\ }\href
  {https://doi.org/10.1103/PhysRevD.31.725} {\bibfield  {journal} {\bibinfo
  {journal} {Phys. Rev. D}\ }\textbf {\bibinfo {volume} {31}},\ \bibinfo
  {pages} {725} (\bibinfo {year} {1985})}\BibitemShut {NoStop}%
\bibitem [{\citenamefont {Muronga}(2002)}]{Muronga:2001zk}%
  \BibitemOpen
  \bibfield  {author} {\bibinfo {author} {\bibfnamefont {A.}~\bibnamefont
  {Muronga}},\ }\bibfield  {title} {\bibinfo {title} {{Second order dissipative
  fluid dynamics for ultrarelativistic nuclear collisions}},\ }\href
  {https://doi.org/10.1103/PhysRevLett.88.062302} {\bibfield  {journal}
  {\bibinfo  {journal} {Phys. Rev. Lett.}\ }\textbf {\bibinfo {volume} {88}},\
  \bibinfo {pages} {062302} (\bibinfo {year} {2002})},\ \bibinfo {note}
  {[Erratum: Phys.Rev.Lett. 89, 159901 (2002)]},\ \Eprint
  {https://arxiv.org/abs/nucl-th/0104064} {arXiv:nucl-th/0104064} \BibitemShut
  {NoStop}%
\bibitem [{\citenamefont {Rocha}\ \emph {et~al.}(2024)\citenamefont {Rocha},
  \citenamefont {Wagner}, \citenamefont {Denicol}, \citenamefont {Noronha},\
  and\ \citenamefont {Rischke}}]{Rocha:2023ilf}%
  \BibitemOpen
  \bibfield  {author} {\bibinfo {author} {\bibfnamefont {G.~S.}\ \bibnamefont
  {Rocha}}, \bibinfo {author} {\bibfnamefont {D.}~\bibnamefont {Wagner}},
  \bibinfo {author} {\bibfnamefont {G.~S.}\ \bibnamefont {Denicol}}, \bibinfo
  {author} {\bibfnamefont {J.}~\bibnamefont {Noronha}},\ and\ \bibinfo {author}
  {\bibfnamefont {D.~H.}\ \bibnamefont {Rischke}},\ }\bibfield  {title}
  {\bibinfo {title} {{Theories of Relativistic Dissipative Fluid Dynamics}},\
  }\href {https://doi.org/10.3390/e26030189} {\bibfield  {journal} {\bibinfo
  {journal} {Entropy}\ }\textbf {\bibinfo {volume} {26}},\ \bibinfo {pages}
  {189} (\bibinfo {year} {2024})},\ \Eprint {https://arxiv.org/abs/2311.15063}
  {arXiv:2311.15063 [nucl-th]} \BibitemShut {NoStop}%
\bibitem [{\citenamefont {Bjorken}(1983)}]{Bjorken:1982qr}%
  \BibitemOpen
  \bibfield  {author} {\bibinfo {author} {\bibfnamefont {J.~D.}\ \bibnamefont
  {Bjorken}},\ }\bibfield  {title} {\bibinfo {title} {{Highly Relativistic
  Nucleus-Nucleus Collisions: The Central Rapidity Region}},\ }\href
  {https://doi.org/10.1103/PhysRevD.27.140} {\bibfield  {journal} {\bibinfo
  {journal} {Phys. Rev. D}\ }\textbf {\bibinfo {volume} {27}},\ \bibinfo
  {pages} {140} (\bibinfo {year} {1983})}\BibitemShut {NoStop}%
\bibitem [{\citenamefont {Gubser}(2010)}]{Gubser:2010ze}%
  \BibitemOpen
  \bibfield  {author} {\bibinfo {author} {\bibfnamefont {S.~S.}\ \bibnamefont
  {Gubser}},\ }\bibfield  {title} {\bibinfo {title} {{Symmetry constraints on
  generalizations of Bjorken flow}},\ }\href
  {https://doi.org/10.1103/PhysRevD.82.085027} {\bibfield  {journal} {\bibinfo
  {journal} {Phys. Rev. D}\ }\textbf {\bibinfo {volume} {82}},\ \bibinfo
  {pages} {085027} (\bibinfo {year} {2010})},\ \Eprint
  {https://arxiv.org/abs/1006.0006} {arXiv:1006.0006 [hep-th]} \BibitemShut
  {NoStop}%
\bibitem [{\citenamefont {Torrieri}\ \emph {et~al.}(2008)\citenamefont
  {Torrieri}, \citenamefont {Tomasik},\ and\ \citenamefont
  {Mishustin}}]{Torrieri:2007fb}%
  \BibitemOpen
  \bibfield  {author} {\bibinfo {author} {\bibfnamefont {G.}~\bibnamefont
  {Torrieri}}, \bibinfo {author} {\bibfnamefont {B.}~\bibnamefont {Tomasik}},\
  and\ \bibinfo {author} {\bibfnamefont {I.}~\bibnamefont {Mishustin}},\
  }\bibfield  {title} {\bibinfo {title} {{Bulk Viscosity driven clusterization
  of quark-gluon plasma and early freeze-out in relativistic heavy-ion
  collisions}},\ }\href {https://doi.org/10.1103/PhysRevC.77.034903} {\bibfield
   {journal} {\bibinfo  {journal} {Phys. Rev. C}\ }\textbf {\bibinfo {volume}
  {77}},\ \bibinfo {pages} {034903} (\bibinfo {year} {2008})},\ \Eprint
  {https://arxiv.org/abs/0707.4405} {arXiv:0707.4405 [nucl-th]} \BibitemShut
  {NoStop}%
\bibitem [{\citenamefont {Rajagopal}\ and\ \citenamefont
  {Tripuraneni}(2010)}]{Rajagopal:2009yw}%
  \BibitemOpen
  \bibfield  {author} {\bibinfo {author} {\bibfnamefont {K.}~\bibnamefont
  {Rajagopal}}\ and\ \bibinfo {author} {\bibfnamefont {N.}~\bibnamefont
  {Tripuraneni}},\ }\bibfield  {title} {\bibinfo {title} {{Bulk Viscosity and
  Cavitation in Boost-Invariant Hydrodynamic Expansion}},\ }\href
  {https://doi.org/10.1007/JHEP03(2010)018} {\bibfield  {journal} {\bibinfo
  {journal} {JHEP}\ }\textbf {\bibinfo {volume} {03}},\ \bibinfo {pages}
  {018}},\ \Eprint {https://arxiv.org/abs/0908.1785} {arXiv:0908.1785 [hep-ph]}
  \BibitemShut {NoStop}%
\bibitem [{\citenamefont {Bhatt}\ \emph
  {et~al.}(2010{\natexlab{a}})\citenamefont {Bhatt}, \citenamefont {Mishra},\
  and\ \citenamefont {Sreekanth}}]{Bhatt:2010cy}%
  \BibitemOpen
  \bibfield  {author} {\bibinfo {author} {\bibfnamefont {J.~R.}\ \bibnamefont
  {Bhatt}}, \bibinfo {author} {\bibfnamefont {H.}~\bibnamefont {Mishra}},\ and\
  \bibinfo {author} {\bibfnamefont {V.}~\bibnamefont {Sreekanth}},\ }\bibfield
  {title} {\bibinfo {title} {{Thermal photons in QGP and non-ideal effects}},\
  }\href {https://doi.org/10.1007/JHEP11(2010)106} {\bibfield  {journal}
  {\bibinfo  {journal} {JHEP}\ }\textbf {\bibinfo {volume} {11}},\ \bibinfo
  {pages} {106}},\ \Eprint {https://arxiv.org/abs/1011.1969} {arXiv:1011.1969
  [hep-ph]} \BibitemShut {NoStop}%
\bibitem [{\citenamefont {Bhatt}\ \emph
  {et~al.}(2010{\natexlab{b}})\citenamefont {Bhatt}, \citenamefont {Mishra},\
  and\ \citenamefont {Sreekanth}}]{Bhatt:2010hu}%
  \BibitemOpen
  \bibfield  {author} {\bibinfo {author} {\bibfnamefont {J.~R.}\ \bibnamefont
  {Bhatt}}, \bibinfo {author} {\bibfnamefont {H.}~\bibnamefont {Mishra}},\ and\
  \bibinfo {author} {\bibfnamefont {V.}~\bibnamefont {Sreekanth}},\ }\bibfield
  {title} {\bibinfo {title} {{Cavitation and thermal photon production in
  relativistic heavy ion collisions}},\ }\href@noop {} {\  (\bibinfo {year}
  {2010}{\natexlab{b}})},\ \Eprint {https://arxiv.org/abs/1005.2756}
  {arXiv:1005.2756 [hep-ph]} \BibitemShut {NoStop}%
\bibitem [{\citenamefont {Bhatt}\ \emph {et~al.}(2012)\citenamefont {Bhatt},
  \citenamefont {Mishra},\ and\ \citenamefont {Sreekanth}}]{Bhatt:2011kx}%
  \BibitemOpen
  \bibfield  {author} {\bibinfo {author} {\bibfnamefont {J.~R.}\ \bibnamefont
  {Bhatt}}, \bibinfo {author} {\bibfnamefont {H.}~\bibnamefont {Mishra}},\ and\
  \bibinfo {author} {\bibfnamefont {V.}~\bibnamefont {Sreekanth}},\ }\bibfield
  {title} {\bibinfo {title} {{Cavitation and thermal dilepton production in
  QGP}},\ }\href {https://doi.org/10.1016/j.nuclphysa.2011.11.012} {\bibfield
  {journal} {\bibinfo  {journal} {Nucl. Phys. A}\ }\textbf {\bibinfo {volume}
  {875}},\ \bibinfo {pages} {181} (\bibinfo {year} {2012})},\ \Eprint
  {https://arxiv.org/abs/1101.5597} {arXiv:1101.5597 [hep-ph]} \BibitemShut
  {NoStop}%
\bibitem [{\citenamefont {Habich}\ and\ \citenamefont
  {Romatschke}(2014)}]{Habich:2014tpa}%
  \BibitemOpen
  \bibfield  {author} {\bibinfo {author} {\bibfnamefont {M.}~\bibnamefont
  {Habich}}\ and\ \bibinfo {author} {\bibfnamefont {P.}~\bibnamefont
  {Romatschke}},\ }\bibfield  {title} {\bibinfo {title} {{Onset of cavitation
  in the quark-gluon plasma}},\ }\href
  {https://doi.org/10.1007/JHEP12(2014)054} {\bibfield  {journal} {\bibinfo
  {journal} {JHEP}\ }\textbf {\bibinfo {volume} {12}},\ \bibinfo {pages}
  {054}},\ \Eprint {https://arxiv.org/abs/1405.1978} {arXiv:1405.1978 [hep-ph]}
  \BibitemShut {NoStop}%
\bibitem [{\citenamefont {Byres}\ \emph {et~al.}(2020)\citenamefont {Byres},
  \citenamefont {Lim}, \citenamefont {McGinn}, \citenamefont {Ouellette},\ and\
  \citenamefont {Nagle}}]{Byres:2019xld}%
  \BibitemOpen
  \bibfield  {author} {\bibinfo {author} {\bibfnamefont {M.}~\bibnamefont
  {Byres}}, \bibinfo {author} {\bibfnamefont {S.~H.}\ \bibnamefont {Lim}},
  \bibinfo {author} {\bibfnamefont {C.}~\bibnamefont {McGinn}}, \bibinfo
  {author} {\bibfnamefont {J.}~\bibnamefont {Ouellette}},\ and\ \bibinfo
  {author} {\bibfnamefont {J.~L.}\ \bibnamefont {Nagle}},\ }\bibfield  {title}
  {\bibinfo {title} {{Bulk viscosity and cavitation in heavy ion collisions}},\
  }\href {https://doi.org/10.1103/PhysRevC.101.044902} {\bibfield  {journal}
  {\bibinfo  {journal} {Phys. Rev. C}\ }\textbf {\bibinfo {volume} {101}},\
  \bibinfo {pages} {044902} (\bibinfo {year} {2020})},\ \Eprint
  {https://arxiv.org/abs/1910.12930} {arXiv:1910.12930 [nucl-th]} \BibitemShut
  {NoStop}%
\bibitem [{\citenamefont {Naik}\ and\ \citenamefont
  {Sreekanth}(2023)}]{Naik:2022pyk}%
  \BibitemOpen
  \bibfield  {author} {\bibinfo {author} {\bibfnamefont {L.~J.}\ \bibnamefont
  {Naik}}\ and\ \bibinfo {author} {\bibfnamefont {V.}~\bibnamefont
  {Sreekanth}},\ }\bibfield  {title} {\bibinfo {title} {{Second order
  hydrodynamics based on effective kinetic theory and electromagnetic signals
  from QGP}},\ }\href {https://doi.org/10.1088/1361-6471/aca924} {\bibfield
  {journal} {\bibinfo  {journal} {J. Phys. G}\ }\textbf {\bibinfo {volume}
  {50}},\ \bibinfo {pages} {025102} (\bibinfo {year} {2023})},\ \Eprint
  {https://arxiv.org/abs/2207.05310} {arXiv:2207.05310 [nucl-th]} \BibitemShut
  {NoStop}%
\bibitem [{\citenamefont {Bhatt}\ \emph {et~al.}(2011)\citenamefont {Bhatt},
  \citenamefont {Mishra},\ and\ \citenamefont {Sreekanth}}]{Bhatt:2011kr}%
  \BibitemOpen
  \bibfield  {author} {\bibinfo {author} {\bibfnamefont {J.~R.}\ \bibnamefont
  {Bhatt}}, \bibinfo {author} {\bibfnamefont {H.}~\bibnamefont {Mishra}},\ and\
  \bibinfo {author} {\bibfnamefont {V.}~\bibnamefont {Sreekanth}},\ }\bibfield
  {title} {\bibinfo {title} {{Shear viscosity, cavitation and hydrodynamics at
  LHC}},\ }\href {https://doi.org/10.1016/j.physletb.2011.09.052} {\bibfield
  {journal} {\bibinfo  {journal} {Phys. Lett. B}\ }\textbf {\bibinfo {volume}
  {704}},\ \bibinfo {pages} {486} (\bibinfo {year} {2011})},\ \Eprint
  {https://arxiv.org/abs/1103.4333} {arXiv:1103.4333 [hep-ph]} \BibitemShut
  {NoStop}%
\bibitem [{\citenamefont {Matsui}\ \emph {et~al.}(1986)\citenamefont {Matsui},
  \citenamefont {Svetitsky},\ and\ \citenamefont {McLerran}}]{Matsui:1985eu}%
  \BibitemOpen
  \bibfield  {author} {\bibinfo {author} {\bibfnamefont {T.}~\bibnamefont
  {Matsui}}, \bibinfo {author} {\bibfnamefont {B.}~\bibnamefont {Svetitsky}},\
  and\ \bibinfo {author} {\bibfnamefont {L.~D.}\ \bibnamefont {McLerran}},\
  }\bibfield  {title} {\bibinfo {title} {{Strangeness Production in
  Ultrarelativistic Heavy Ion Collisions. 1. Chemical Kinetics in the Quark -
  Gluon Plasma}},\ }\href {https://doi.org/10.1103/PhysRevD.37.844} {\bibfield
  {journal} {\bibinfo  {journal} {Phys. Rev. D}\ }\textbf {\bibinfo {volume}
  {34}},\ \bibinfo {pages} {783} (\bibinfo {year} {1986})},\ \bibinfo {note}
  {[Erratum: Phys.Rev.D 37, 844 (1988)]}\BibitemShut {NoStop}%
\bibitem [{\citenamefont {Biro}\ \emph {et~al.}(1993)\citenamefont {Biro},
  \citenamefont {van Doorn}, \citenamefont {Muller}, \citenamefont {Thoma},\
  and\ \citenamefont {Wang}}]{Biro:1993qt}%
  \BibitemOpen
  \bibfield  {author} {\bibinfo {author} {\bibfnamefont {T.~S.}\ \bibnamefont
  {Biro}}, \bibinfo {author} {\bibfnamefont {E.}~\bibnamefont {van Doorn}},
  \bibinfo {author} {\bibfnamefont {B.}~\bibnamefont {Muller}}, \bibinfo
  {author} {\bibfnamefont {M.~H.}\ \bibnamefont {Thoma}},\ and\ \bibinfo
  {author} {\bibfnamefont {X.~N.}\ \bibnamefont {Wang}},\ }\bibfield  {title}
  {\bibinfo {title} {{Parton equilibration in relativistic heavy ion
  collisions}},\ }\href {https://doi.org/10.1103/PhysRevC.48.1275} {\bibfield
  {journal} {\bibinfo  {journal} {Phys. Rev. C}\ }\textbf {\bibinfo {volume}
  {48}},\ \bibinfo {pages} {1275} (\bibinfo {year} {1993})},\ \Eprint
  {https://arxiv.org/abs/nucl-th/9303004} {arXiv:nucl-th/9303004} \BibitemShut
  {NoStop}%
\bibitem [{\citenamefont {Levai}\ \emph {et~al.}(1995)\citenamefont {Levai},
  \citenamefont {Muller},\ and\ \citenamefont {Wang}}]{Levai:1994dx}%
  \BibitemOpen
  \bibfield  {author} {\bibinfo {author} {\bibfnamefont {P.}~\bibnamefont
  {Levai}}, \bibinfo {author} {\bibfnamefont {B.}~\bibnamefont {Muller}},\ and\
  \bibinfo {author} {\bibfnamefont {X.-N.}\ \bibnamefont {Wang}},\ }\bibfield
  {title} {\bibinfo {title} {{Open charm production in an equilibrating parton
  plasma}},\ }\href {https://doi.org/10.1103/PhysRevC.51.3326} {\bibfield
  {journal} {\bibinfo  {journal} {Phys. Rev. C}\ }\textbf {\bibinfo {volume}
  {51}},\ \bibinfo {pages} {3326} (\bibinfo {year} {1995})},\ \Eprint
  {https://arxiv.org/abs/hep-ph/9412352} {arXiv:hep-ph/9412352} \BibitemShut
  {NoStop}%
\bibitem [{\citenamefont {Baier}\ \emph {et~al.}(2001)\citenamefont {Baier},
  \citenamefont {Mueller}, \citenamefont {Schiff},\ and\ \citenamefont
  {Son}}]{Baier:2000sb}%
  \BibitemOpen
  \bibfield  {author} {\bibinfo {author} {\bibfnamefont {R.}~\bibnamefont
  {Baier}}, \bibinfo {author} {\bibfnamefont {A.~H.}\ \bibnamefont {Mueller}},
  \bibinfo {author} {\bibfnamefont {D.}~\bibnamefont {Schiff}},\ and\ \bibinfo
  {author} {\bibfnamefont {D.~T.}\ \bibnamefont {Son}},\ }\bibfield  {title}
  {\bibinfo {title} {{'Bottom up' thermalization in heavy ion collisions}},\
  }\href {https://doi.org/10.1016/S0370-2693(01)00191-5} {\bibfield  {journal}
  {\bibinfo  {journal} {Phys. Lett. B}\ }\textbf {\bibinfo {volume} {502}},\
  \bibinfo {pages} {51} (\bibinfo {year} {2001})},\ \Eprint
  {https://arxiv.org/abs/hep-ph/0009237} {arXiv:hep-ph/0009237} \BibitemShut
  {NoStop}%
\bibitem [{\citenamefont {Berges}\ \emph {et~al.}(2014)\citenamefont {Berges},
  \citenamefont {Boguslavski}, \citenamefont {Schlichting},\ and\ \citenamefont
  {Venugopalan}}]{Berges:2013eia}%
  \BibitemOpen
  \bibfield  {author} {\bibinfo {author} {\bibfnamefont {J.}~\bibnamefont
  {Berges}}, \bibinfo {author} {\bibfnamefont {K.}~\bibnamefont {Boguslavski}},
  \bibinfo {author} {\bibfnamefont {S.}~\bibnamefont {Schlichting}},\ and\
  \bibinfo {author} {\bibfnamefont {R.}~\bibnamefont {Venugopalan}},\
  }\bibfield  {title} {\bibinfo {title} {{Turbulent thermalization process in
  heavy-ion collisions at ultrarelativistic energies}},\ }\href
  {https://doi.org/10.1103/PhysRevD.89.074011} {\bibfield  {journal} {\bibinfo
  {journal} {Phys. Rev. D}\ }\textbf {\bibinfo {volume} {89}},\ \bibinfo
  {pages} {074011} (\bibinfo {year} {2014})},\ \Eprint
  {https://arxiv.org/abs/1303.5650} {arXiv:1303.5650 [hep-ph]} \BibitemShut
  {NoStop}%
\bibitem [{\citenamefont {Kurkela}\ and\ \citenamefont
  {Mazeliauskas}(2019{\natexlab{a}})}]{Kurkela:2018oqw}%
  \BibitemOpen
  \bibfield  {author} {\bibinfo {author} {\bibfnamefont {A.}~\bibnamefont
  {Kurkela}}\ and\ \bibinfo {author} {\bibfnamefont {A.}~\bibnamefont
  {Mazeliauskas}},\ }\bibfield  {title} {\bibinfo {title} {{Chemical
  equilibration in weakly coupled QCD}},\ }\href
  {https://doi.org/10.1103/PhysRevD.99.054018} {\bibfield  {journal} {\bibinfo
  {journal} {Phys. Rev. D}\ }\textbf {\bibinfo {volume} {99}},\ \bibinfo
  {pages} {054018} (\bibinfo {year} {2019}{\natexlab{a}})},\ \Eprint
  {https://arxiv.org/abs/1811.03068} {arXiv:1811.03068 [hep-ph]} \BibitemShut
  {NoStop}%
\bibitem [{\citenamefont {Kurkela}\ \emph {et~al.}(2019)\citenamefont
  {Kurkela}, \citenamefont {Mazeliauskas}, \citenamefont {Paquet},
  \citenamefont {Schlichting},\ and\ \citenamefont {Teaney}}]{Kurkela:2018vqr}%
  \BibitemOpen
  \bibfield  {author} {\bibinfo {author} {\bibfnamefont {A.}~\bibnamefont
  {Kurkela}}, \bibinfo {author} {\bibfnamefont {A.}~\bibnamefont
  {Mazeliauskas}}, \bibinfo {author} {\bibfnamefont {J.-F.}\ \bibnamefont
  {Paquet}}, \bibinfo {author} {\bibfnamefont {S.}~\bibnamefont
  {Schlichting}},\ and\ \bibinfo {author} {\bibfnamefont {D.}~\bibnamefont
  {Teaney}},\ }\bibfield  {title} {\bibinfo {title} {{Effective kinetic
  description of event-by-event pre-equilibrium dynamics in high-energy
  heavy-ion collisions}},\ }\href {https://doi.org/10.1103/PhysRevC.99.034910}
  {\bibfield  {journal} {\bibinfo  {journal} {Phys. Rev. C}\ }\textbf {\bibinfo
  {volume} {99}},\ \bibinfo {pages} {034910} (\bibinfo {year} {2019})},\
  \Eprint {https://arxiv.org/abs/1805.00961} {arXiv:1805.00961 [hep-ph]}
  \BibitemShut {NoStop}%
\bibitem [{\citenamefont {Kurkela}\ and\ \citenamefont
  {Mazeliauskas}(2019{\natexlab{b}})}]{Kurkela:2018xxd}%
  \BibitemOpen
  \bibfield  {author} {\bibinfo {author} {\bibfnamefont {A.}~\bibnamefont
  {Kurkela}}\ and\ \bibinfo {author} {\bibfnamefont {A.}~\bibnamefont
  {Mazeliauskas}},\ }\bibfield  {title} {\bibinfo {title} {{Chemical
  Equilibration in Hadronic Collisions}},\ }\href
  {https://doi.org/10.1103/PhysRevLett.122.142301} {\bibfield  {journal}
  {\bibinfo  {journal} {Phys. Rev. Lett.}\ }\textbf {\bibinfo {volume} {122}},\
  \bibinfo {pages} {142301} (\bibinfo {year} {2019}{\natexlab{b}})},\ \Eprint
  {https://arxiv.org/abs/1811.03040} {arXiv:1811.03040 [hep-ph]} \BibitemShut
  {NoStop}%
\bibitem [{\citenamefont {El}\ \emph {et~al.}(2009)\citenamefont {El},
  \citenamefont {Muronga}, \citenamefont {Xu},\ and\ \citenamefont
  {Greiner}}]{El:2008yy}%
  \BibitemOpen
  \bibfield  {author} {\bibinfo {author} {\bibfnamefont {A.}~\bibnamefont
  {El}}, \bibinfo {author} {\bibfnamefont {A.}~\bibnamefont {Muronga}},
  \bibinfo {author} {\bibfnamefont {Z.}~\bibnamefont {Xu}},\ and\ \bibinfo
  {author} {\bibfnamefont {C.}~\bibnamefont {Greiner}},\ }\bibfield  {title}
  {\bibinfo {title} {{Shear viscosity and out of equilibrium dissipative
  hydrodynamics}},\ }\href {https://doi.org/10.1103/PhysRevC.79.044914}
  {\bibfield  {journal} {\bibinfo  {journal} {Phys. Rev. C}\ }\textbf {\bibinfo
  {volume} {79}},\ \bibinfo {pages} {044914} (\bibinfo {year} {2009})},\
  \Eprint {https://arxiv.org/abs/0812.2762} {arXiv:0812.2762 [hep-ph]}
  \BibitemShut {NoStop}%
\bibitem [{\citenamefont {Bhatt}\ and\ \citenamefont
  {Sreekanth}(2010)}]{Bhatt:2009zg}%
  \BibitemOpen
  \bibfield  {author} {\bibinfo {author} {\bibfnamefont {J.~R.}\ \bibnamefont
  {Bhatt}}\ and\ \bibinfo {author} {\bibfnamefont {V.}~\bibnamefont
  {Sreekanth}},\ }\bibfield  {title} {\bibinfo {title} {{Photon emission from
  out of equilibrium dissipative parton plasma}},\ }\href
  {https://doi.org/10.1142/S0218301310014765} {\bibfield  {journal} {\bibinfo
  {journal} {Int. J. Mod. Phys. E}\ }\textbf {\bibinfo {volume} {19}},\
  \bibinfo {pages} {299} (\bibinfo {year} {2010})},\ \Eprint
  {https://arxiv.org/abs/0901.1363} {arXiv:0901.1363 [hep-ph]} \BibitemShut
  {NoStop}%
\bibitem [{\citenamefont {Naik}\ and\ \citenamefont
  {Sreekanth}(2026)}]{Naik:2026ehp}%
  \BibitemOpen
  \bibfield  {author} {\bibinfo {author} {\bibfnamefont {L.~J.}\ \bibnamefont
  {Naik}}\ and\ \bibinfo {author} {\bibfnamefont {V.}~\bibnamefont
  {Sreekanth}},\ }\bibfield  {title} {\bibinfo {title} {{Finite-density
  dynamics of chemically equilibrating QGP in conformal Gubser flow and hard
  thermal photon production}},\ }\href
  {https://doi.org/10.1088/1361-6471/ae8463} {\bibfield  {journal} {\bibinfo
  {journal} {J. Phys. G}\ }\textbf {\bibinfo {volume} {53}},\ \bibinfo {pages}
  {075106} (\bibinfo {year} {2026})},\ \Eprint
  {https://arxiv.org/abs/2606.31749} {arXiv:2606.31749 [hep-ph]} \BibitemShut
  {NoStop}%
\bibitem [{\citenamefont {Denicol}\ \emph {et~al.}(2010)\citenamefont
  {Denicol}, \citenamefont {Koide},\ and\ \citenamefont
  {Rischke}}]{Denicol:2010xn}%
  \BibitemOpen
  \bibfield  {author} {\bibinfo {author} {\bibfnamefont {G.~S.}\ \bibnamefont
  {Denicol}}, \bibinfo {author} {\bibfnamefont {T.}~\bibnamefont {Koide}},\
  and\ \bibinfo {author} {\bibfnamefont {D.~H.}\ \bibnamefont {Rischke}},\
  }\bibfield  {title} {\bibinfo {title} {{Dissipative relativistic fluid
  dynamics: a new way to derive the equations of motion from kinetic theory}},\
  }\href {https://doi.org/10.1103/PhysRevLett.105.162501} {\bibfield  {journal}
  {\bibinfo  {journal} {Phys. Rev. Lett.}\ }\textbf {\bibinfo {volume} {105}},\
  \bibinfo {pages} {162501} (\bibinfo {year} {2010})},\ \Eprint
  {https://arxiv.org/abs/1004.5013} {arXiv:1004.5013 [nucl-th]} \BibitemShut
  {NoStop}%
\end{thebibliography}%

\end{document}